\documentclass[aip, pof, graphicx, 12pt, floatfix, preprint]{revtex4-1}

\usepackage{subcaption}
\usepackage{graphicx}
\usepackage{dcolumn}
\usepackage{multirow}
\usepackage{ctable}
\usepackage{booktabs}
\usepackage{amsmath}
\usepackage{amssymb}
\usepackage{xcolor}
\usepackage{makecell}
\usepackage{comment}
\usepackage{hyperref}
\usepackage[dvipsnames]{xcolor}
\hypersetup{hidelinks}
\setcitestyle{numbers,square}
\makeatletter\AtBeginDocument{\def\place@bibnumber#1{[#1]\hskip0.5em}}\makeatother

\definecolor{g}{rgb}{0., 0.4, 0.1}
\newcommand{\hide}[1]{}

\begin{document}

\title{A Particle Multi-Relaxation Bhatnagar-Gross-Krook Method for Rarefied Monatomic Gas Mixtures}

\author{Inchan Kim}
\affiliation{Department of Aerospace Engineering, Korea Advanced Institute of Science and Technology, Daejeon, 34141, Republic of Korea}

\author{Joonbeom Kim}
\affiliation{Department of Aerospace Engineering, Korea Advanced Institute of Science and Technology, Daejeon, 34141, Republic of Korea}

\author{Woonghwi Park}
\affiliation{Department of Aerospace Engineering, Korea Advanced Institute of Science and Technology, Daejeon, 34141, Republic of Korea}

\author{Eunji Jun}
\email{eunji.jun@kaist.ac.kr}
\affiliation{Department of Aerospace Engineering, Korea Advanced Institute of Science and Technology, Daejeon, 34141, Republic of Korea}


\begin{abstract}

Kinetic models based on the Bhatnagar–Gross–Krook (BGK) framework provide an efficient alternative to the Boltzmann equation for rarefied gas flows; however, existing formulations for gas mixtures remain limited in representing pair-dependent relaxation processes and recovering correct Navier–Stokes–Fourier (NSF) transport behavior. A particle-based unified BGK (UBGK) model for monatomic gas mixtures is developed by extending the single-species UBGK framework to a multi-relaxation formulation. The model preserves the pairwise interaction structure of the mixture Boltzmann equation, enabling independent species-pair relaxations for an arbitrary number of species. The relaxation properties of the mixture UBGK model are determined by matching the production terms to those of the Boltzmann equation, ensuring correct NSF-level transport behavior. The model is implemented within the particle framework and validated against DSMC using four benchmark cases: homogeneous relaxation, Poiseuille flow, Couette flow, and hypersonic flow around a cylinder. The results demonstrate that the mixture UBGK model captures species-specific non-equilibrium effects, including species-dependent differences in velocity and temperature, across a range of mole fractions and Knudsen numbers in good agreement with DSMC. Furthermore, cost and accuracy analyses show that the mixture UBGK model becomes more efficient than DSMC at sufficiently large time step sizes, but its first-order accuracy suggests further improvement through higher-order schemes.

\end{abstract}


\maketitle 

\section{\label{sec:intro}Introduction}


Multi-scale gas flows arise in a wide range of aerospace applications, including high-altitude flight and hypersonic reentry. In such flows, the degree of rarefaction varies across the domain, and continuum and non-equilibrium regimes may coexist. To simulate such flows, the Direct Simulation Monte Carlo (DSMC) method is widely used. DSMC models gas flows using simulated particles that undergo advection and collisions, in accordance with the Boltzmann equation.\cite{Bird1994-DSMC} However, the computational cost of DSMC increases significantly in the near-continuum regimes, where the decreasing mean collision time and mean free path tighten temporal and spatial resolution requirements.\cite{Gallis2009-DSMCConstraints} To overcome this computational overhead, alternative approaches such as kinetic models\cite{Jun2019-FPSlitOrifice, Fei2020-BGKShockwave, Kim2024-FPDSMCHybrid} have been developed. In particular, the Bhatnagar-Gross-Krook (BGK) models\cite{yao2023-SBGK+, Gao2014-BGKCompare} and the Fokker–Planck (FP) models\cite{Jun2018-FPDSMC, Kim2022-VHSFP, Kim2024-SOFP, Kim2024-FPInterpolation} approximate the Boltzmann equation while preserving conservation laws and near-continuum transport properties, enabling more efficient particle-based simulations in near-continuum and multi-scale applications.\cite{Pfeiffer2018-BGKComparison}


BGK models approximate intermolecular collisions by a relaxation process, where the particle distribution function relaxes toward the target function. In the standard BGK model,\cite{Bhatnagar1954-BGK} the target function is described by a Maxwellian, and the model satisfies the laws of conservation. However, it yields an incorrect Prandtl number ($\mathrm{Pr}$ = 1), whereas the correct value for monatomic gases is $2/3$. This is because shear stress and heat flux cannot be independently modified with a single relaxation rate. To recover the correct $\mathrm{Pr}$, modified BGK models, such as the ellipsoidal statistical BGK (ESBGK) model\cite{Holway1966-ESBGK, Cercignani1988-ESBGK}, the Shakhov BGK (SBGK) model\cite{Shakhov1968-SBGK}, and the unified BGK (UBGK) model\cite{Chen2015-UBGK} have been introduced. The ESBGK and SBGK models fix $\mathrm{Pr}$ through unique mechanisms. While the ESBGK model modifies shear stress by employing an anisotropic Gaussian as the target function, the SBGK model modifies the heat flux by adapting a skewed Maxwellian. From a mathematical standpoint, the ESBGK model globally fulfills the H-theorem, whereas the SBGK model satisfies it only for near-equilibrium conditions.\cite{Andries2000-ESBGKHtheorem, Shakhov1968-SBGKHtheorem} From a numerical standpoint, however, several shock-wave problems have demonstrated that the SBGK model provides more accurate predictions of temperature and number density profiles in the shock region than the ESBGK model.\cite{Fei2020-BGKShockwave, Park2024-BGKCompare} To address these limitations, the UBGK model is constructed to combine the ESBGK model and the SBGK model. $\mathrm{Pr}$ is fixed by introducing two auxiliary coefficients that control the relative contribution of shear stress and heat flux modifications. Although the UBGK model satisfies the H-theorem only in near-equilibrium conditions, the additional degree of freedom provided by the auxiliary coefficients allows improved prediction of temperature and number density distributions compared to both ESBGK and SBGK models.\cite{Chen2015-UBGK} Following the development of particle-based BGK methods for monatomic gases,\cite{Gallis2000-ParticleBGK, Macrossan2001-ParticleBGK} BGK models have been extended to diatomic BGK models,\cite{Pfeiffer2018-DiatomicBGK, Mathiaud2022-DiatomicBGK} and polyatomic BGK models.\cite{Pfeiffer2019-PolyatomicBGK, Fei2022-PolyatomicBGK}


Extending BGK models to gas mixtures has become an active area of research in recent years.\cite{Pirner2018-BGKMixtureCat} In gas mixtures, transport processes involve not only viscosity and thermal conductivity, but also mass diffusion and thermal diffusion, which together constitute the NSF-level transport behavior.\cite{Bird1994-DSMC} Thus, $\mathrm{Pr}$ alone, which relates viscosity and thermal conductivity, is insufficient to fully characterize transport in gas mixtures. Existing BGK models for gas mixtures can be categorized into two formulations: single-relaxation and multi-relaxation.\cite{Pirner2018-BGKMixtureCat} The single-relaxation BGK formulation, which was established by Andries \textit{et al.},\cite{Andries2002-SRMixBGK} employs a single effective relaxation by collapsing self-species and cross-species interactions into a single relaxation per species. However, the Andries model cannot fully recover correct transport coefficients, because the target distribution is described with a Maxwellian. Subsequent developments have modified the Andries model into mixture ESBGK and SBGK models by introducing adjustable parameters to recover correct transport coefficients.\cite{Brull2015-SRMixBGK, Groppi2011-SRMixBGK, Todorova2019-SRMixBGK, Pfeiffer2021-SRMixBGK} Owing to the compact structure, single-relaxation models have been widely adopted, and have demonstrated acceptable accuracy in many applications.\cite{Todorova2020-SRDiMixBGK, Brull2021-SRDiMixBGK, Bisi2018-SRMixBGKCR, Hild2024-SRPolyMixBGK, Fei2026-SRPolyMixBGK} However, because each species uses a single relaxation rate, the single-relaxation models' capability of capturing species-pair dependent relaxation processes is limited. Multi-relaxation formulations, on the other hand, adopt multiple relaxation terms whose mathematical structure resembles that of the mixture Boltzmann equation, thereby allowing pair-dependent relaxations. Early works of Gross \& Krook\cite{Gross1956-MRMixBGK} established the multi-relaxation BGK formulation. Since then, progress has been made in improving mathematical consistency with the structure of the Boltzmann equation, as well as enforcement of conservation laws and the H-theorem.\cite{Morse1964-MRMixBGK, Hamel1965-MRMixBGK, Greene1973-MRMixBGK, Haack2017-MRMixBGK, Klingenberg2017-MRMixBGK, Bobylev2018-MRMixBGK} Despite these efforts, the target distribution remained as a Maxwellian, due to the increased mathematical complexity and difficulty in determining model parameters associated with multi-species interactions. As a result, {accurately reproducing  mixture transport phenomena remained not fully resolved. The SBGK-based mixture model proposed by Li \textit{et al.}\cite{Li2024-MRMixBGK} is, to the author's knowledge, the only multi-relaxation model that correctly reproduces all transport coefficients. However, the model derivation relies on the Maxwell intermolecular potential, and is limited to binary gas mixtures. Taken together, these observations suggest that a mixture BGK model that simultaneously preserves the mathematical structure of the mixture Boltzmann equation, is applicable to an arbitrary number of species, and reproduces correct NSF-level transport behavior has yet to be established.


In this work, a new multi-relaxation mixture UBGK model for monatomic gases is developed, which is applicable to an arbitrary number of species, and reproduces correct NSF-level transport behavior. The single-species UBGK model is extended to a multi-relaxation mixture formulation, and accurate transport behavior in the near-continuum regime is recovered by matching the production terms with those of the Boltzmann equation. A stochastic algorithm for particle-based simulation is presented, and the accuracy of the model is assessed through representative test cases. The remainder of this paper is as follows: In Section \ref{sec:ubgk-model}, kinetic theory for monatomic gas mixtures is reviewed. The construction of the mixture UBGK model is introduced, and the associated relaxation properties are derived by matching the production terms with those of the Boltzmann equation. Section \ref{sec:numericalimplementation} describes the numerical algorithm for particle-based simulation. Section \ref{sec:results} presents the numerical results of four test cases: homogeneous relaxation, Poiseuille flow, Couette flow, and hypersonic flow around a cylinder. Furthermore, computational efficiency and accuracy of the mixture UBGK model compared to DSMC are examined. Finally, Section \ref{sec:conclusions} provides a summary and concludes the paper.

\newpage
\section{\label{sec:ubgk-model}Unified BGK model for monatomic gas mixtures}

\subsection{\label{subsec:kinetic} Kinetic theory and macroscopic properties of gas mixtures}


The state of a gas is described statistically through the velocity distribution function (VDF), $f^\alpha \left( \boldsymbol{c}^\alpha, \boldsymbol{x}, t \right)$.\cite{Bird1994-DSMC} The quantity $f^\alpha \left( \boldsymbol{c}^\alpha, \boldsymbol{x}, t \right)d\boldsymbol{c}^\alpha d\boldsymbol{x}$ gives the number of molecules of species $\alpha$ that occupy the phase-space volume $d\boldsymbol{c}^\alpha d\boldsymbol{x}$ around velocity $\boldsymbol{c}^\alpha$ and location $\boldsymbol{x}$, at time $t$. Throughout this work, the Greek superscript is used to denote the species. In the absence of external forces, the evolution of the VDF of each species in an $N_S$-component mixture is governed by a system of Boltzmann equations, given by
\begin{equation}\label{eqn:Boltzmann equation}
\begin{aligned}
    \frac{\partial{f^\alpha}}{\partial{t}} + c_i \frac{\partial{f^\alpha}}{\partial{x_i}} = \sum_{\beta=1}^{N_S} S^{\alpha\beta},
\end{aligned}
\end{equation}
where the collision term for species $\alpha$ is decomposed into pairwise collision terms for an arbitrary colliding pair ($\alpha - \beta$), denoted by $S^{\alpha\beta}$. The corresponding pairwise collision term can be described by the Boltzmann collision operator:
\begin{equation}\label{eqn:Boltzmann collision operator}
\begin{aligned}
    S_{Boltz}^{\alpha\beta} = \int_{-\infty}^{\infty} \int_{0}^{4 \pi} \Big( f'^\alpha f'^\beta - f^\alpha f^\beta \Big) c_r^{\alpha\beta} \sigma^{\alpha\beta} d\Omega d\boldsymbol{c}^\beta,
\end{aligned}
\end{equation}
where ($f^\alpha$, $f^\beta$) and ($f'^\alpha$, $f'^\beta$) are the pre-collision and post-collision VDFs, $c_r^{\alpha\beta} = \left| \boldsymbol{c}^\beta - \boldsymbol{c}^\alpha \right|$ is the magnitude of relative velocity between two colliding molecules, $\sigma^{\alpha\beta}$ is the collision differential cross section, and $\Omega$ denotes the solid angle. The integrals over $d\Omega$ and $d\boldsymbol{c}^\beta$ account for all post-collision scattering directions and velocities.


The macroscopic properties of species $\alpha$ can be obtained by evaluating velocity moments of the VDF:
\begin{equation}\label{eqn:massrho-species}
\begin{aligned}
    \rho^\alpha = m^\alpha n^\alpha = m^\alpha \int_{-\infty}^{\infty} {f^\alpha d\boldsymbol{c}^\alpha},
\end{aligned}
\end{equation}
\begin{equation}\label{eqn:velocity-species}
\begin{aligned}
    u_i^\alpha = \frac{1}{n^\alpha} \int_{-\infty}^{\infty} {c_i^\alpha f^\alpha d\boldsymbol{c}^\alpha},
\end{aligned}
\end{equation}
\begin{equation}\label{eqn:temperature-species}
\begin{aligned}
    \frac{3}{2} k_B T^\alpha = \frac{1}{n^\alpha} \int_{-\infty}^{\infty} {\frac{1}{2} m^\alpha C_j^\alpha C_j^\alpha f^\alpha d\boldsymbol{c}^\alpha},
\end{aligned}
\end{equation}
\begin{equation}\label{eqn:pressure-species}
\begin{aligned}
    p_{ij}^\alpha = \int_{-\infty}^{\infty} {m^\alpha C_i^\alpha C_j^\alpha f^\alpha d\boldsymbol{c}^\alpha},
\end{aligned}
\end{equation}
\begin{equation}\label{eqn:shearstress-species}
\begin{aligned}
    \sigma_{ij}^\alpha = p_{\langle ij \rangle}^\alpha = \int_{-\infty}^{\infty} {m^\alpha C_{\langle i}^\alpha C_{j \rangle}^\alpha f^\alpha d\boldsymbol{c}^\alpha},
\end{aligned}
\end{equation}
\begin{equation}\label{eqn:heatflux-species}
\begin{aligned}
    q_i^\alpha = \int_{-\infty}^{\infty} {\frac{1}{2} m^\alpha C_i^\alpha C_j^\alpha C_j^\alpha f^\alpha d\boldsymbol{c}^\alpha},
\end{aligned}
\end{equation}
where $\rho^\alpha$ is the mass density,  $m^\alpha$ is the molecular mass, $n^\alpha$ is the number density, $u_i^\alpha$ is the velocity, $T^\alpha$ is the temperature, $p_{ij}^\alpha$ is the pressure tensor, $\sigma_{ij}^\alpha$ is the shear stress tensor, $q_i^\alpha$ is the heat flux, $c_i^\alpha$ is the molecular velocity, $C_i^\alpha = c_i^\alpha - u_i^\alpha$ is the thermal velocity, and $k_B$ is the Boltzmann constant. Throughout this work, the Einstein summation convention is adopted for all repeated indices, and the angle brackets $\langle \cdot \rangle$ denote the deviatoric component of a symmetric tensor. The total number density, total mass density, and bulk velocity of the whole mixture can be obtained as: 
\begin{equation}\label{eqn:nrho-mixture}
\begin{aligned}
    n = \sum_{\alpha=1}^{N_S} n^\alpha
\end{aligned}
\end{equation}
\begin{equation}\label{eqn:massnrho-mixture}
\begin{aligned}
    \rho = \sum_{\alpha=1}^{N_S} n^\alpha m^\alpha
\end{aligned}
\end{equation}
\begin{equation}\label{eqn:velocity-mixture}
\begin{aligned}
    v_i = \frac{1}{\rho} \sum_{\alpha=1}^{N_S} \rho^\alpha u_i^\alpha
\end{aligned}
\end{equation}


An alternative approach for evaluating macroscopic moments is to use the thermal velocity defined relative to the mixture bulk velocity, rather than the species bulk velocity, i.e., $\hat{C}_i^\alpha = c_i^\alpha - v_i$.\cite{Gupta2015-BoltzmannPT, Kim2025-DiMixFP} With this definition, the corresponding properties of species $\alpha$ become:
\begin{equation}\label{eqn:velocityhat-species}
\begin{aligned}
    \hat{u}_i^\alpha = \frac{1}{n^\alpha} \int_{-\infty}^{\infty} {\hat{C}_i^\alpha f^\alpha d\boldsymbol{c}^\alpha} = u_i^\alpha - v_i,
\end{aligned}
\end{equation}
\begin{equation}\label{eqn:temperaturehat-species}
\begin{aligned}
    \frac{3}{2} k_B \hat{T}^\alpha = \frac{1}{n^\alpha} \int_{-\infty}^{\infty}{\frac{1}{2} m^\alpha \hat{C}_j^\alpha \hat{C}_j^\alpha f^\alpha d\boldsymbol{c}^\alpha},
\end{aligned}
\end{equation}
\begin{equation}\label{eqn:pressurehat-species}
\begin{aligned}
    \hat{p}_{ij}^\alpha = \int_{-\infty}^{\infty} {m^\alpha \hat{C}_i^\alpha \hat{C}_j^\alpha f^\alpha d\boldsymbol{c}^\alpha},
\end{aligned}
\end{equation}
\begin{equation}\label{eqn:shearstresshat-species}
\begin{aligned}
    \hat{\sigma}_{ij}^\alpha = \hat{p}_{\langle ij \rangle}^\alpha = \int_{-\infty}^{\infty} {m^\alpha \hat{C}_{\langle i}^\alpha \hat{C}_{j \rangle}^\alpha f^\alpha d\boldsymbol{c}^\alpha},
\end{aligned}
\end{equation}
\begin{equation}\label{eqn:heatfluxhat-species}
\begin{aligned}
    \hat{q}_i^\alpha = \int_{-\infty}^{\infty}{\frac{1}{2} m^\alpha \hat{C}_i^\alpha \hat{C}_j^\alpha \hat{C}_j^\alpha f^\alpha d\boldsymbol{c}^\alpha},
\end{aligned}
\end{equation}
where $\hat{u}_i^\alpha$ is commonly termed as the diffusion velocity. Using the definition of hat properties, the temperature, pressure tensor, shear stress tensor, and heat flux of the whole mixture can be obtained as:
\begin{equation}\label{eqn:temperaturehat-mixture}
\begin{aligned}
    T = \frac{1}{n} \sum_{\alpha = 1}^{N_S} n^\alpha \hat{T}^\alpha,
\end{aligned}
\end{equation}
\begin{equation}\label{eqn:pressurehat-mixture}
\begin{aligned}
    p_{ij} = \sum_{\alpha = 1}^{N_S} \hat{p}_{ij}^\alpha,
\end{aligned}
\end{equation}
\begin{equation}\label{eqn:shearstresshat-mixture}
\begin{aligned}
    \sigma_{ij} = \sum_{\alpha = 1}^{N_S} \hat{\sigma}_{ij}^\alpha,
\end{aligned}
\end{equation}
\begin{equation}\label{eqn:heatfluxhat-mixture}
\begin{aligned}
    q_i = \sum_{\alpha = 1}^{N_S} \hat{q}_i^\alpha.
\end{aligned}
\end{equation}

\subsection{\label{subsec:BoltzmannProductionTerms} Boltzmann production terms}


The production terms of the Boltzmann equation are obtained by taking velocity moments of the Boltzmann collision operator:\cite{Gupta2012-BoltzmannCI}
\begin{equation}\label{eqn:Boltzmann PT}
\begin{aligned}
    P_{Boltz}^{\alpha\beta} \left( \psi_i^\alpha \right) = \int_{- \infty}^{\infty} m^\alpha \psi_i^\alpha S_{Boltz}^{\alpha\beta} d\boldsymbol{c}^\alpha,
\end{aligned}
\end{equation}
where $\psi_i^\alpha$ denotes a velocity polynomial. In this work, the set of velocity polynomials considered is restricted to $\psi_i^\alpha \in \{1, \allowbreak \hat{C}_i^\alpha, \allowbreak \hat{C}_j^\alpha \hat{C}_j^\alpha, \allowbreak \hat{C}_{\langle i}^\alpha \hat{C}_{j \rangle}^\alpha, \allowbreak \hat{C}_i^\alpha \hat{C}_j^\alpha \hat{C}_j^\alpha\}$, which corresponds to mass, momentum, energy, shear stress, and heat flux.\cite{Gupta2015-BoltzmannPT}


Gupta \& Torrilhon\cite{Gupta2012-BoltzmannCI, Gupta2015-BoltzmannPT} evaluated the Boltzmann production terms using the Grad's 13-moment function.\cite{Struchtrup2005-Grad13} Originally developed for Maxwell and hard sphere molecular models, this formulation was later extended to variable hard sphere (VHS) models by Hepp \textit{et al.}\cite{Hepp2020-FPMixVHS} and Kim \& Jun.\cite{Kim2025-FPMix, Kim2025-DiMixFP} In the present work, the formulation of Kim \& Jun is adopted. However, instead of using the modified collision rate introduced therein, the collision rate derived directly from the Boltzmann collision operator is employed for consistency. The resulting VHS-scaled Boltzmann production terms are expressed as follows:
\begin{equation}\label{eqn:G13 PT-mass}
\begin{aligned}
    P_{G13}^{\alpha\beta} \left( 1 \right) = 0,
\end{aligned}
\end{equation}
\begin{equation}\label{eqn:G13 PT-momentum}
\begin{aligned}
    P_{G13}^{\alpha\beta} \left( \hat{C}_i^\alpha \right)
    &= -\nu_{G13}^{\alpha\beta} \mu^\beta \rho^\alpha \Bigg[ 
        \frac{5}{3} \mathrm{VHS} \left[ 1 \right] \left( \hat{u}_i^\alpha - \hat{u}_i^\beta \right)
        + \frac{1}{6 \hat{\theta}^{\alpha\beta}} \mathrm{VHS} \left[ 2 \right] \left( \frac{\hat{h}_i^\alpha}{\rho^\alpha} - \frac{\hat{h}_i^\beta}{\rho^\beta} \right)
    \Bigg],
\end{aligned}
\end{equation}
\begin{equation}\label{eqn:G13 PT-energy}
\begin{aligned}
    P_{G13}^{\alpha\beta} \left( \hat{C}_j^\alpha \hat{C}_j^\alpha \right)
    &= -\nu_{G13}^{\alpha\beta} \mu^\beta \rho^\alpha \mathrm{VHS} \left[ 1 \right] \Bigg[ 
        10 \hat{\theta}^{\alpha\beta} \Delta \hat{\theta}^{\alpha\beta} 
        -\frac{10}{3} \left( \mu^\alpha - \mu^\beta \right) \hat{u}_j^\alpha \hat{u}_j^\beta
    \Bigg],
\end{aligned}
\end{equation}
\begin{equation}\label{eqn:G13 PT-shearstress}
\begin{aligned}
    P_{G13}^{\alpha\beta} \left( \hat{C}_{\langle i}^\alpha \hat{C}_{j \rangle}^\alpha \right)
    &= -\nu_{G13}^{\alpha\beta} \mu^\beta \rho^\alpha \Bigg[
        4 \mu^\beta \bigg\{
            \frac{\hat{\sigma}_{ij}}{\rho^\alpha}
            + \frac{1}{3} \mathrm{VHS}[3] \left( \frac{\hat{\sigma}_{ij}^\alpha}{\rho^\alpha} - \frac{\hat{\sigma}_{ij}^\beta}{\rho^\beta} \right) 
        \bigg\} \\
        &\quad + \frac{10}{3} \mathrm{VHS}[1] (\mu^\alpha - \mu^\beta) \frac{\hat{\sigma}_{ij}^\alpha}{\rho^\alpha} 
        + \frac{1}{3} \Delta \hat{\theta}^{\alpha\beta} \left( \frac{\hat{\sigma}_{ij}^\alpha}{\rho^\alpha} +  \frac{\hat{\sigma}_{ij}^\beta}{\rho^\beta} \right)
    \Bigg],
\end{aligned}
\end{equation}
\begin{equation}\label{eqn:G13 PT-heatflux}
\begin{aligned}
    P_{G13}^{\alpha\beta} \left( \hat{C}_i^\alpha \hat{C}_j^\alpha \hat{C}_j^\alpha \right)
    &= -\nu_{G13}^{\alpha\beta} \mu^\beta \rho^\alpha \Bigg[
        \frac{16}{3} \mu^\beta \frac{\hat{h}_i^\alpha}{\rho^\alpha}
        + 10 \mathrm{VHS} \left[ 1 \right] \left( \mu^\alpha - \mu^\beta \right) \frac{\hat{h}_i^\alpha}{\rho^\alpha} \\
        &\quad + \frac{5 + 27 \mu^\beta}{3} \mu^\beta \mathrm{VHS} \left[ 4 \right] \left( \frac{\hat{h}_i^\alpha}{\rho^\alpha} - \frac{\hat{h}_i^\beta}{\rho^\beta} \right) \\
        &\quad + 10 \frac{5 + \mu^\beta}{3} \hat{\theta}^{\alpha\beta} \mu^\beta \mathrm{VHS} \left[ 5 \right] \left( \hat{u}_i^\alpha - \hat{u}_i^\beta \right) \\
        &\quad + \frac{\Delta \hat{\theta}^{\alpha\beta}}{2} \bigg\{
            \left( 9 - 8 \mu^\beta - \frac{1}{2} \Delta \hat{\theta}^{\alpha\beta} \right) \frac{\hat{h}_i^\alpha}{\rho^\alpha} \\
            &\quad\quad - \frac{1}{3} \left( 5 - 24 \mu^\beta - \frac{3}{2} \Delta \hat{\theta}^{\alpha\beta} \right) \frac{\hat{h}_i^\beta}{\rho^\beta}
        \bigg\} \\
        &\quad + 5 \Delta \hat{\theta}^{\alpha\beta} \bigg\{
            \left( 5 - 4 \mu^\beta + \frac{1}{2} \Delta \hat{\theta}^{\alpha\beta} \right) \hat{\theta}^{\alpha\beta} \hat{u}_i^\alpha \\
            &\quad\quad - \frac{1}{3} \left( 5 - 12 \mu^\beta + \frac{3}{2} \Delta \hat{\theta}^{\alpha\beta} \right) \hat{\theta}^{\alpha\beta} \hat{u}_i^\beta
        \bigg\}
    \Bigg],
\end{aligned}
\end{equation}
where the collision rate and the reduced properties are given as:
\begin{equation}\label{eqn:G13 PT-collisionrate}
\begin{aligned}
    \nu_{G13}^{\alpha\beta}
    &= \frac{16}{5} \mathrm{VHS} \left[ 6 \right] \sqrt{\pi} n^\beta \left( d_{ref}^{\alpha\beta} \right)^2 \sqrt{\hat{\theta}^{\alpha\beta}},
\end{aligned}
\end{equation}
and 
\begin{equation}\label{eqn:reduced-massfraction}
\begin{aligned}
    \mu^\beta
    &= \frac{m^\beta}{m^\alpha + m^\beta},
\end{aligned}
\end{equation}
\begin{equation}\label{eqn:reduced-temperature}
\begin{aligned}
    \hat{\theta}^\alpha
    &= R^\alpha \hat{T}^\alpha = \frac{k_B}{m^\alpha}\hat{T}^\alpha,
\end{aligned}
\end{equation}
\begin{equation}\label{eqn:reduced-temperatures}
\begin{aligned}
    \hat{\theta}^{\alpha\beta}
    &= \frac{\hat{\theta}^\alpha + \hat{\theta}^\beta}{2},
\end{aligned}
\end{equation}
\begin{equation}\label{eqn:reduced-temperaturediff}
\begin{aligned}
    \Delta \hat{\theta}^{\alpha\beta}
    &= \frac{\mu^\alpha \hat{\theta}^\alpha - \mu^\beta \hat{\theta}^\beta}{\hat{\theta}^{\alpha\beta}},
\end{aligned}
\end{equation}
\begin{equation}\label{eqn:reduced-heatflux}
\begin{aligned}
    \hat{h}_i^\alpha = \hat{q}^\alpha - \frac{5}{2} \hat{p}^\alpha \hat{u}_i^\alpha.
\end{aligned}
\end{equation}
Moreover, the VHS-scaling parameters are given as:
\begin{equation}\label{eqn:VHSscaling-1}
\begin{aligned}
    \mathrm{VHS} \left[ 1 \right]
    &= 3 \frac{\Gamma \left( 3.5 - \omega^{\alpha\beta} \right)}{\Gamma \left( 4.5 - \omega^{\alpha\beta} \right)},
\end{aligned}
\end{equation}
\begin{equation}\label{eqn:VHSscaling-2}
\begin{aligned}
    \mathrm{VHS} \left[ 2 \right]
    &= 6 - 15\frac{\Gamma \left( 3.5 - \omega^{\alpha\beta} \right)}{\Gamma \left( 4.5 - \omega^{\alpha\beta} \right)},
\end{aligned}
\end{equation}
\begin{equation}\label{eqn:VHSscaling-3}
\begin{aligned}
    \mathrm{VHS} \left[ 3 \right]
    &= \frac{15}{2} \frac{\Gamma \left( 3.5 - \omega^{\alpha\beta} \right)}{\Gamma \left( 4.5 - \omega^{\alpha\beta} \right)} - \frac{3}{2},
\end{aligned}
\end{equation}
\begin{equation}\label{eqn:VHSscaling-4}
\begin{aligned}
    \mathrm{VHS} \left[ 4 \right]
    &= - \frac{1}{5 + 27 \mu^\beta} \Bigg[ 
        10 \mu^\alpha \left( \frac{15}{2} \frac{\Gamma \left( 3.5 - \omega^{\alpha\beta} \right)}{\Gamma \left( 4.5 - \omega^{\alpha\beta} \right)} - 3 \right) \\
        &\quad + 4 \mu^\beta \left( \frac{23}{2} + \frac{45}{2} \frac{\Gamma \left( 3.5 - \omega^{\alpha\beta} \right)}{\Gamma \left( 4.5 - \omega^{\alpha\beta} \right)} - 3 \frac{\Gamma \left( 5.5 - \omega^{\alpha\beta} \right)}{\Gamma \left( 4.5 - \omega^{\alpha\beta} \right)} \right)
    \Bigg],
\end{aligned}
\end{equation}
\begin{equation}\label{eqn:VHSscaling-5}
\begin{aligned}
    \mathrm{VHS} \left[ 5 \right]
    &= \frac{1}{5 + \mu^\beta} \Bigg[
        15 \mu^\alpha \frac{\Gamma \left( 3.5 - \omega^{\alpha\beta} \right)}{\Gamma \left( 4.5 - \omega^{\alpha\beta} \right)}  + 6 \mu^\beta
    \Bigg],
\end{aligned}
\end{equation}
\begin{equation}\label{eqn:VHSscaling-6}
\begin{aligned}
    \mathrm{VHS} \left[ 6 \right]
    &= \frac{1}{6} \frac{\Gamma \left( 4.5 - \omega^{\alpha\beta} \right)}{\Gamma \left( 2.5 - \omega^{\alpha\beta} \right)} \left( \frac{k_B T_{ref}^{\alpha\beta}}{2~m_r^{\alpha\beta} \hat{\theta}^{\alpha\beta}} \right)^{\omega^{\alpha\beta} - 0.5},
\end{aligned}
\end{equation}
where $\Gamma \left( \cdot \right)$ denotes the standard Gamma function, $\omega^{\alpha\beta}$ is the VHS viscosity exponent, $T_{ref}^{\alpha\beta}$ is the reference temperature, and $m_r^{\alpha\beta} = m^\alpha m^\beta/ \left( m^\alpha + m^\beta \right)$ is the reduced mass. Equations (\ref{eqn:G13 PT-mass})$-$(\ref{eqn:G13 PT-heatflux}) are referred to as the G13 production terms, and Eq. (\ref{eqn:G13 PT-collisionrate}) is referred to as the G13 collision rate for this work.

\subsection{\label{subsec:BGKPT}The multi-relaxation BGK equation for gas mixtures and its production terms}


A multi-relaxation BGK formulation for gas mixtures, applicable to an arbitrary number of species, is adopted to approximate the Boltzmann collision operator while preserving its pairwise interaction structure. Following the framework of Bobylev \textit{et al.},\cite{Bobylev2018-MRMixBGK}, the total collision term is decomposed into pairwise relaxation terms. The corresponding BGK relaxation term for an arbitrary relaxing pair ($\alpha - \beta$) is written as:
\begin{equation}\label{eqn:BGK relaxation operator}
\begin{aligned}
    S_{BGK}^{\alpha\beta} = \nu_{BGK}^{\alpha\beta} \left( f_t^{\alpha\beta} - f^\alpha \right),
\end{aligned}
\end{equation}
where $f_t^{\alpha\beta}$ is the target distribution function for a relaxing pair, and $\nu_{BGK}^{\alpha\beta}$ is the corresponding relaxation rate.


For the BGK model to be valid in the near-continuum regime, it has to satisfy the conservation laws of mass, momentum, and energy, and recover correct NSF-level transport.\cite{Kim2025-FPMix} These requirements are satisfied by ensuring that the model reproduces the Boltzmann production terms, which reads:
\begin{equation}\label{eqn:Boltzmann PT-BGK PT}
\begin{aligned}
    P_{Boltz}^{\alpha\beta} \left( \psi_i^\alpha \right) = P_{BGK}^{\alpha\beta} \left( \psi_i^\alpha \right),
\end{aligned}
\end{equation}
where the production terms of the BGK equation are obtained with:
\begin{equation}\label{eqn:BGK PT}
\begin{aligned}
    P_{BGK}^{\alpha\beta} \left( \psi_i^\alpha \right) = \int_{- \infty}^{\infty} m^\alpha \psi_i^\alpha S_{BGK}^{\alpha\beta} d\boldsymbol{c}^\alpha.
\end{aligned}
\end{equation}
By substituting Eq. (\ref{eqn:BGK relaxation operator}) into Eq. (\ref{eqn:BGK PT}), one can obtain the following expression:
\begin{equation}\label{eqn:BGK PT-derivation}
\begin{aligned}
    P_{BGK}^{\alpha\beta} \left( \psi_i^\alpha \right) 
    &= \int_{-\infty}^{\infty} m^\alpha \psi_i^\alpha \nu_{BGK}^{\alpha\beta} \left( f_t^{\alpha\beta} - f^\alpha \right) d\boldsymbol{c}^\alpha.
\end{aligned}
\end{equation}
Consequently, the remaining task for the mixture BGK model construction is to implement an appropriate target distribution function and a relaxation rate.

\subsection{\label{subsec:UBGK} The Unified BGK mixture model}


A new mixture UBGK model for an arbitrary number of species is introduced by using the UBGK framework. For a relaxing pair $(\alpha - \beta)$, the target distribution is written as:
\begin{equation}\label{eqn:UBGK-TargetDistribution}
\begin{aligned}
    f_U^{\alpha\beta} = G'^{\alpha\beta} + S'^{\alpha\beta}
\end{aligned}
\end{equation}
\begin{subequations}
\begin{equation}\label{eqn:UBGK-TargetDistribution_ES1}
\begin{aligned}
    G'^{\alpha\beta} = \frac{n^\alpha}{\left( \mathrm{det} \left| 2 \pi \lambda'^{\alpha\beta} \right| \right)^{1/2}} \mathrm{exp} \left( -\frac{1}{2} C_i^{\alpha\beta} \left(\lambda'^{\alpha\beta} \right)^{-1} C_j^{\alpha\beta} \right)
\end{aligned}
\end{equation}
\begin{equation}\label{eqn:UBGK-TargetDistribution_ES2}
\begin{aligned}
    \lambda'^{\alpha\beta} = R^\alpha T^{\alpha\beta} \delta_{ij} + C_{ES} \frac{\sigma_{ij}^{\alpha\beta}}{\rho^\alpha}
\end{aligned}
\end{equation}
\end{subequations}
\begin{equation}\label{eqn:UBGK-TargetDistribution_S}
\begin{aligned}
    S'^{\alpha\beta} = G'^{\alpha\beta} \Bigg[ 
        C_S \frac{2 C_j^{\alpha\beta} q_j^{\alpha\beta}}{5 \rho^\alpha \left( R^\alpha T^{\alpha\beta} \right)^2 } \left( \frac{ C_j^{\alpha\beta} C_j^{\alpha\beta} }{ 2 R^\alpha T^{\alpha\beta} } - \frac{5}{2} \right)
    \Bigg],
\end{aligned}
\end{equation}
where $G'^{\alpha\beta}$ and $S'^{\alpha\beta}$ correspond to the ESBGK and SBGK distributions, respectively, $\lambda'^{\alpha\beta}$ is the anisotropic stress tensor, and $C_i^{\alpha\beta}=c_i^\alpha-u^{\alpha\beta}$ is the thermal velocity with respect to the relaxation velocity. For the unification of ESBGK and SBGK models, auxiliary parameters $C_{ES}$ and $C_S$ were introduced to achieve correct $\mathrm{Pr}$.\cite{Chen2015-UBGK} The parameters, namely $u_i^{\alpha\beta}, \allowbreak T^{\alpha\beta}, \allowbreak \sigma_{ij}^{\alpha\beta}$, and $q_i^{\alpha\beta}$, are the relaxation properties that constitute the UBGK model's target distribution function. To ensure positive definiteness of the anisotropic stress tensor $\lambda'^{\alpha\beta}$, $C_{ES}$ must be taken between the range $[-1/2, 1)$.\cite{Chen2015-UBGK, Pfeiffer2018-BGKComparison, Mathiaud2016-ESFP} With these auxiliary coefficients, the resulting Prandtl number for the UBGK model is $Pr = \frac{1 - C_S}{1 - C_{ES}}$.


A suitable relaxation rate must be specified. Under near-equilibrium conditions, i.e., $\hat{T}^\alpha \approx \hat{T}^\beta$, the G13 collision rate from Eq. (\ref{eqn:G13 PT-collisionrate}) is proportional to $\sqrt{1 / m_r^{\alpha\beta}}$. This corresponds to a pair-dependent mass scaling that is symmetric between species. However, since the induced relaxation of macroscopic properties is generally species-dependent, a pair-symmetric collision rate is not suitable. To account for this asymmetry, a correction factor $\sqrt{2 m_r^{\alpha\beta} / m^\alpha}$ is introduced to modify the G13 collision rate, yielding:
\begin{equation}\label{eqn:UBGK-relaxationrate}
\begin{aligned}
    \nu_{UBGK}^{\alpha\beta} = \nu_{G13}^{\alpha\beta} \sqrt{ \frac{2 m_r^{\alpha\beta}}{m^\alpha} }.
\end{aligned}
\end{equation}
With this modification, the UBGK relaxation rate becomes species-dependent, proportional to $\sqrt{1 / m^\alpha}$. Therefore, lighter species exhibit larger relaxation rates than heavier species, which is consistent with the expected physical behavior of gas mixtures.\cite{Li2024-MRMixBGK}


The production terms of the UBGK equation are obtained by substituting Eq. (\ref{eqn:UBGK-TargetDistribution}) into Eq. (\ref{eqn:BGK PT-derivation}):
\begin{equation}\label{eqn:UBGK PT-mass}
\begin{aligned}
    P_{UBGK}^{\alpha\beta} \left( 1 \right)
    &= 0,
\end{aligned}
\end{equation}
\begin{equation}\label{eqn:UBGK PT-momentum}
\begin{aligned}
    P_{UBGK}^{\alpha\beta} \left( \hat{C}_i^\alpha \right)
    &= \nu_{UBGK}^{\alpha\beta} \Big[
        \rho^\alpha \left( u_i^{\alpha\beta} - u_i^\alpha \right)
    \Big],
\end{aligned}
\end{equation}
\begin{equation}\label{eqn:UBGK PT-energy}
\begin{aligned}
    P_{UBGK}^{\alpha\beta} \left( \hat{C}_j^\alpha \hat{C}_j^\alpha \right)
    &= \nu_{UBGK}^{\alpha\beta} \Big[
        3 \rho^\alpha R^\alpha \left( T^{\alpha\beta} - T^\alpha \right)
        + \rho^\alpha \left( \Delta \tilde{u}_j^{\alpha\beta} \Delta \tilde{u}_j^{\alpha\beta} - \hat{u}_j^\alpha \hat{u}_j^\alpha \right)
    \Big],
\end{aligned}
\end{equation}
\begin{equation}\label{eqn:UBGK PT-shearstress}
\begin{aligned}
    P_{UBGK}^{\alpha\beta} \left( \hat{C}_{\langle i}^\alpha \hat{C}_{j \rangle}^\alpha\right)
    &= \nu_{UBGK}^{\alpha\beta}\Big[
        C_{ES} \sigma_{ij}^{\alpha\beta} - \sigma_{ij}^\alpha
        + \rho^\alpha \left( \Delta \tilde{u}_{\langle i}^{\alpha\beta} \Delta \tilde{u}_{j \rangle}^{\alpha\beta} - \hat{u}_{\langle i}^\alpha \hat{u}_{j \rangle}^\alpha \right)
    \Big],
\end{aligned}
\end{equation}
\begin{equation}\label{eqn:UBGK PT-heatflux}
\begin{aligned}
    P_{UBGK}^{\alpha\beta} \left( \hat{C}_i^\alpha \hat{C}_j^\alpha \hat{C}_j^\alpha \right)
    &= \nu_{UBGK}^{\alpha\beta}\Big[
        2 C_S q_i^{\alpha\beta} - 2 q_i^\alpha
        + 2 C_{ES} \sigma_{ij}^{\alpha\beta} \Delta \tilde{u}_j^{\alpha\beta} - 2 \sigma_{ij}^\alpha \hat{u}_j^\alpha \\
        &\quad + 5 \rho^\alpha R^\alpha \left( \Delta \tilde{u}_i^{\alpha\beta} T^{\alpha\beta} - \hat{u}_i^{\alpha\beta} T^\alpha \right) \\
        &\quad + \rho^\alpha \left( \Delta \tilde{u}_i^{\alpha\beta} \Delta \tilde{u}_j^{\alpha\beta} \Delta \tilde{u}_j^{\alpha\beta} - \hat{u}_i^\alpha \hat{u}_j^\alpha \hat{u}_j^\alpha \right)
    \Big],
\end{aligned}
\end{equation}
and are referred to as the UBGK production terms for this work. Details of the derivation are provided in Appendix \ref{appendix:UBGKPT}. Since both the G13 and UBGK production terms for mass are zero, the relaxation properties, i.e. $u_i^{\alpha\beta}, \allowbreak T^{\alpha\beta}, \allowbreak \sigma_{ij}^{\alpha\beta}$, and $q_i^{\alpha\beta}$, can be explicitly expressed by matching Eqs. (\ref{eqn:UBGK PT-momentum})$-$(\ref{eqn:UBGK PT-heatflux}) with Eqs. (\ref{eqn:G13 PT-momentum})$-$(\ref{eqn:G13 PT-heatflux}):
\begin{equation}\label{eqn:UBGK-relaxation-velocity}
\begin{aligned}
    u_i^{\alpha\beta}
    &= u_i^\alpha \\
    &\quad- \frac{\nu_{G13}^{\alpha\beta}}{\nu_{UBGK}^{\alpha\beta}} \Bigg[
        \frac{5}{3} \mu^\beta \mathrm{VHS} \left[ 1 \right] \left( \hat{u}_i^\alpha - \hat{u}_i^\beta \right)
        + \frac{1}{6 \hat{\theta}^{\alpha\beta}} \mu^\beta \mathrm{VHS} \left[ 2 \right] \left( \frac{\hat{h}_i^\alpha}{\rho^\alpha} - \frac{\hat{h}_i^\beta}{\rho^\beta} \right)
    \Bigg],
\end{aligned}
\end{equation}
\begin{equation}\label{eqn:UBGK-relaxation-temperature}
\begin{aligned}
    T^{\alpha\beta} 
    &= T^\alpha - \frac{1}{3 R^\alpha} \Bigg[
        \Delta \tilde{u}_j^{\alpha\beta} \Delta \tilde{u}_j^{\alpha\beta} - \hat{u}_j^\alpha \hat{u}_j^\alpha \\
        &\quad +\mu^\beta \mathrm{VHS} \left[ 1 \right] \frac{\nu_{G13}^{\alpha\beta}}{\nu_{UBGK}^{\alpha\beta}} \bigg[
            10 \hat{\theta}^{\alpha\beta} \Delta \hat{\theta}^{\alpha\beta} 
            -\frac{10}{3} \left( \mu^\alpha - \mu^\beta \right) \hat{u}_j^\alpha \hat{u}_j^\beta
        \bigg]
    \Bigg],
\end{aligned}
\end{equation}
\begin{equation}\label{eqn:UBGK-relaxation-shearstress}
\begin{aligned}
    \sigma_{ij}^{\alpha\beta}
    &= \frac{1}{C_{ES}} \Bigg[
        \sigma_{ij}^\alpha - \rho^\alpha \left( \Delta \tilde{u}_{\langle i}^{\alpha\beta} \Delta \tilde{u}_{j \rangle}^{\alpha\beta} - \hat{u}_{\langle i}^\alpha \hat{u}_{j \rangle}^\alpha \right) \\
        &\quad - \rho^\alpha \mu^\beta \frac{\nu_{G13}^{\alpha\beta}}{\nu_{UBGK}^{\alpha\beta}} \bigg[
            4 \mu^\beta \bigg\{
                \frac{\hat{\sigma}_{ij}}{\rho^\alpha}
                + \frac{1}{3} \mathrm{VHS}[3] \left( \frac{\hat{\sigma}_{ij}^\alpha}{\rho^\alpha} - \frac{\hat{\sigma}_{ij}^\beta}{\rho^\beta} \right)
            \bigg\} \\
            &\quad + \frac{10}{3} \mathrm{VHS}[1] (\mu^\alpha - \mu^\beta) \frac{\hat{\sigma}_{ij}^\alpha}{\rho^\alpha}
            + \frac{1}{3} \Delta \hat{\theta}^{\alpha\beta} \left( \frac{\hat{\sigma}_{ij}^\alpha}{\rho^\alpha} +  \frac{\hat{\sigma}_{ij}^\beta}{\rho^\beta} \right)
        \bigg]
    \Bigg],
\end{aligned}
\end{equation}
\begin{equation}\label{eqn:UBGK-relaxation-heatflux}
\begin{aligned}
    q_{i}^{\alpha\beta}
    &= \frac{1}{C_S} \Bigg[
        q_i^\alpha - C_{ES} \sigma_{ij}^{\alpha\beta} \Delta \tilde{u}_j^{\alpha\beta} + \sigma_{ij}^\alpha \hat{u}_j^\alpha \\
        &\quad - \frac{5}{2} \rho^\alpha R^\alpha \left( \Delta \tilde{u}_i^{\alpha\beta} T^{\alpha\beta} - \hat{u}_i^\alpha T^\alpha \right) \\
        &\quad - \frac{1}{2} \rho^\alpha \left( \Delta \tilde{u}_i^{\alpha\beta} \Delta \tilde{u}_j^{\alpha\beta} \Delta \tilde{u}_j^{\alpha\beta} - \hat{u}_i^\alpha \hat{u}_j^\alpha \hat{u}_j^\alpha \right) \\
        &\quad - \frac{\rho^\alpha \mu^\beta}{2} \frac{\nu_{G13}^{\alpha\beta}}{\nu_{UBGK}^{\alpha\beta}} \bigg[ 
            \frac{16}{3} \mu^\beta \frac{\hat{h}_i^\alpha}{\rho^\alpha}
            + 10 \mathrm{VHS} \left[ 1 \right] \left( \mu^\alpha - \mu^\beta \right) \frac{\hat{h}_i^\alpha}{\rho^\alpha} \\
            &\quad\quad + \frac{5 + 27 \mu^\beta}{3} \mu^\beta \mathrm{VHS} \left[ 4 \right] \left( \frac{\hat{h}_i^\alpha}{\rho^\alpha} - \frac{\hat{h}_i^\beta}{\rho^\beta} \right)
            + 10 \frac{5 + \mu^\beta}{3} \hat{\theta}^{\alpha\beta} \mu^\beta \mathrm{VHS} \left[ 5 \right] \left( \hat{u}_i^\alpha - \hat{u}_i^\beta \right) \\
            &\quad\quad + \frac{\Delta \hat{\theta}^{\alpha\beta}}{2} \bigg\{
                \left( 9 - 8 \mu^\beta - \frac{1}{2} \Delta \hat{\theta}^{\alpha\beta} \right) \frac{\hat{h}_i^\alpha}{\rho^\alpha}
                - \frac{1}{3} \left( 5 - 24 \mu^\beta - \frac{3}{2} \Delta \hat{\theta}^{\alpha\beta} \right) \frac{\hat{h}_i^\beta}{\rho^\beta}
            \bigg\} \\
            &\quad\quad + 5 \Delta \hat{\theta}^{\alpha\beta} \bigg\{
                \left( 5 - 4 \mu^\beta + \frac{1}{2} \Delta \hat{\theta}^{\alpha\beta} \right) \hat{\theta}^{\alpha\beta} \hat{u}_i^\alpha
                - \frac{1}{3} \left( 5 - 12 \mu^\beta + \frac{3}{2} \Delta \hat{\theta}^{\alpha\beta} \right) \hat{\theta}^{\alpha\beta} \hat{u}_i^\beta
            \bigg\}
        \bigg]
    \Bigg].
\end{aligned}
\end{equation}
It should be noted that the relaxation temperature $T^{\alpha\beta}$ must remain positive for physical admissibility. The corresponding discussion is given in Appendix \ref{appendix:T-sign-Discussion}.

\newpage
\section{\label{sec:numericalimplementation}Numerical implementation}

\subsection{\label{subsec:SolAlg}Solution algorithm of mixture UBGK model}


The particle-based UBGK method de-couples particle advection and relaxation steps.\cite{Fei2020-USPBGK} In the relaxation step, the total relaxation probability that a particle of species $\alpha$ undergoes relaxation is given by:
\begin{equation}\label{eqn:BGK-relaxprobability}
\begin{aligned}
    P_{relax}^\alpha = 1 - \mathrm{exp} \left( - \sum_{\beta = 1}^{N_S} \nu_{UBGK}^{\alpha\beta} \cdot \Delta\mathrm{t} \right),
\end{aligned}
\end{equation}
which accounts for all pairwise interactions between species $\alpha$ and all species in the mixture. If a particle is selected for relaxation, its relaxation partner species must then be determined. Since the total relaxation rate is composed of relaxation rates from each species pair, the partner selection probability that the relaxation occurs through species $\gamma$ is given by:
\begin{equation}\label{eqn:BGK-relaxpartner-probability}
\begin{aligned}
    P \left( \gamma \ |\ \alpha \right) = \frac{\nu_{UBGK}^{\alpha\gamma}}{\sum_{\beta = 1}^{N_S} \nu_{UBGK}^{\alpha\beta}}.
\end{aligned}
\end{equation}
Once the partner species $\gamma$ is selected, the post-relaxation particle velocity is sampled from the corresponding UBGK target distribution $f_U^{\alpha\gamma}$, as defined in Eq. (\ref{eqn:UBGK-TargetDistribution}).


To sample post-relaxation velocities from the UBGK distribution, a direct sampling procedure, constructed by combining the ESBGK model's shear stress modification method with the SBGK model's heat flux correction method, is employed. In contrast to the Metropolis-Hastings method, which was previously used for the single-species UBGK model, the present direct sampling procedure avoids iterative Markov-chain-based sampling and provides a simpler and more efficient velocity sampling scheme for the UBGK model.\cite{Chib1995-MH, Pfeiffer2018-BGKComparison} For a relaxing particle, a three-dimensional random velocity sample $\boldsymbol{\xi}^{(0)} = \left( \xi_x^{(0)}, \xi_y^{(0)}, \xi_z^{(0)} \right)$ is generated from the standard normal distribution. To account for shear stress correction, $\boldsymbol{\xi}^{(0)}$ is transformed using the matrix $S$ as:\cite{Gallis2011-ESBGKSampling}
\begin{equation}\label{eqn:UBGK-eigendecomp1}
\begin{aligned}
    \lambda'^{\alpha\gamma} = VDV^\mathrm{T}, 
\end{aligned}
\end{equation}
\begin{equation}\label{eqn:UBGK-eigendeomp2}
\begin{aligned}
    S = VD^{1/2}V^\mathrm{T}, 
\end{aligned}
\end{equation}
\begin{equation}\label{eqn:UBGK-eigendecomp3}
\begin{aligned}
    \boldsymbol{\xi}^{ES} = S \cdot \boldsymbol{\xi}^{(0)}, 
\end{aligned}
\end{equation}
where $V$ is the matrix of eigenvectors of $\lambda'^{\alpha\gamma}$, $D$ is the diagonal matrix whose components are the corresponding eigenvalues, and $\boldsymbol{\xi}^{ES}$ is the transformed sample. Superscript $\mathrm{T}$ represents the matrix transpose. Next, to account for the heat flux correction, an acceptance-rejection step is applied.\cite{Garcia1998-SBGKSampling} The transformed sample $\boldsymbol{\xi}^{ES}$ is accepted as a valid velocity sample from the UBGK distribution and denoted by $\boldsymbol{\xi}^U$ if
\begin{equation}\label{eqn:UBGK-acceptreject-probability}
\begin{aligned}
    \mathcal{R} < p^{AR}/p_{max}^{AR}, 
\end{aligned}
\end{equation}
where $\mathcal{R}$ is a random number sampled from the standard uniform distribution, and the probability function $p^{AR}$ and the envelope function $p_{max}^{AR}$ are given by:
\begin{equation}\label{eqn:UBGK-acceptreject-probabilityfunction}
\begin{aligned}
    p^{AR} = 1 + C_S \frac{2 \xi_j^{ES} q^{\alpha\gamma}_j}{5 \rho^\alpha ( v_{th}^{\alpha\gamma} )^3} \left( \frac{\xi_j^{ES} \xi_j^{ES}}{2} - \frac{5}{2} \right),
\end{aligned}
\end{equation}
\begin{equation}\label{eqn:UBGK-acceptreject-envelopefunction}
\begin{aligned}
    p_{max}^{AR} = 1 + \frac{8~\mathrm{max}_{k= 1, 2, 3} ( | q^{\alpha\gamma}_k| )}{\rho^\alpha (v_{th}^{\alpha\gamma} )^3},
\end{aligned}
\end{equation}
where the $\mathrm{max} \left( \cdot \right)$ function selects the component of the relaxation heat flux with the largest magnitude, and $v_{th}^{\alpha\gamma} = \sqrt{k_B T^{\alpha\gamma}/m^\alpha}$ is the relaxation thermal velocity associated with the relaxation temperature. Finally, the accepted sample $\boldsymbol{\xi}^U$ is multiplied by the relaxation thermal velocity and is shifted by the relaxation velocity to obtain the post-relaxation particle velocity:
\begin{equation}\label{eqn:UBGK-postrelaxation-velocity}
\begin{aligned}
    c_i^\alpha = u_i^{\alpha\gamma} + \xi_i^U \cdot v_{th}^{\alpha\gamma}.
\end{aligned}
\end{equation}


Due to stochastic sampling procedures during the velocity redistribution step, total momentum and energy are not exactly preserved.\cite{Gallis2011-ESBGKSampling} To remove such numerical errors, momentum and energy conservation after velocity redistribution are enforced at the cell level:
\begin{equation}\label{eqn:MomentumEnergyConservation}
\begin{aligned}
    c_i^{\alpha, (n+1)} = v_i^{(n)} + \left( c_i^{\alpha, (*)} - v_i^{(*)} \right) \sqrt{\frac{E^{(n)}}{E^{(*)}}},
\end{aligned}
\end{equation}
where superscript $(n)$ denotes pre-relaxation state, $(*)$ denotes post-relaxation state, and $(n+1)$ denotes the state after conservation enforcement. The kinetic energy $E$ is computed as:
\begin{equation}\label{eqn:MomentumEnergyConservation-energy}
\begin{aligned}
    E = \sum_{\alpha=1}^{N_S} \sum_{p=1}^{N_P^\alpha} {\frac{1}{2} m^\alpha \hat{C}_{p, j}^\alpha \hat{C}_{p, j}^\alpha},
\end{aligned}
\end{equation}
where $N_P^\alpha$ denotes the number of particles of species $\alpha$, and $\hat{C}_{p, j}^\alpha$ denotes the thermal velocity of particle index $p$ of species $\alpha$.


\begin{figure*}[t] 
    \centering
    \includegraphics[width=0.56\linewidth]{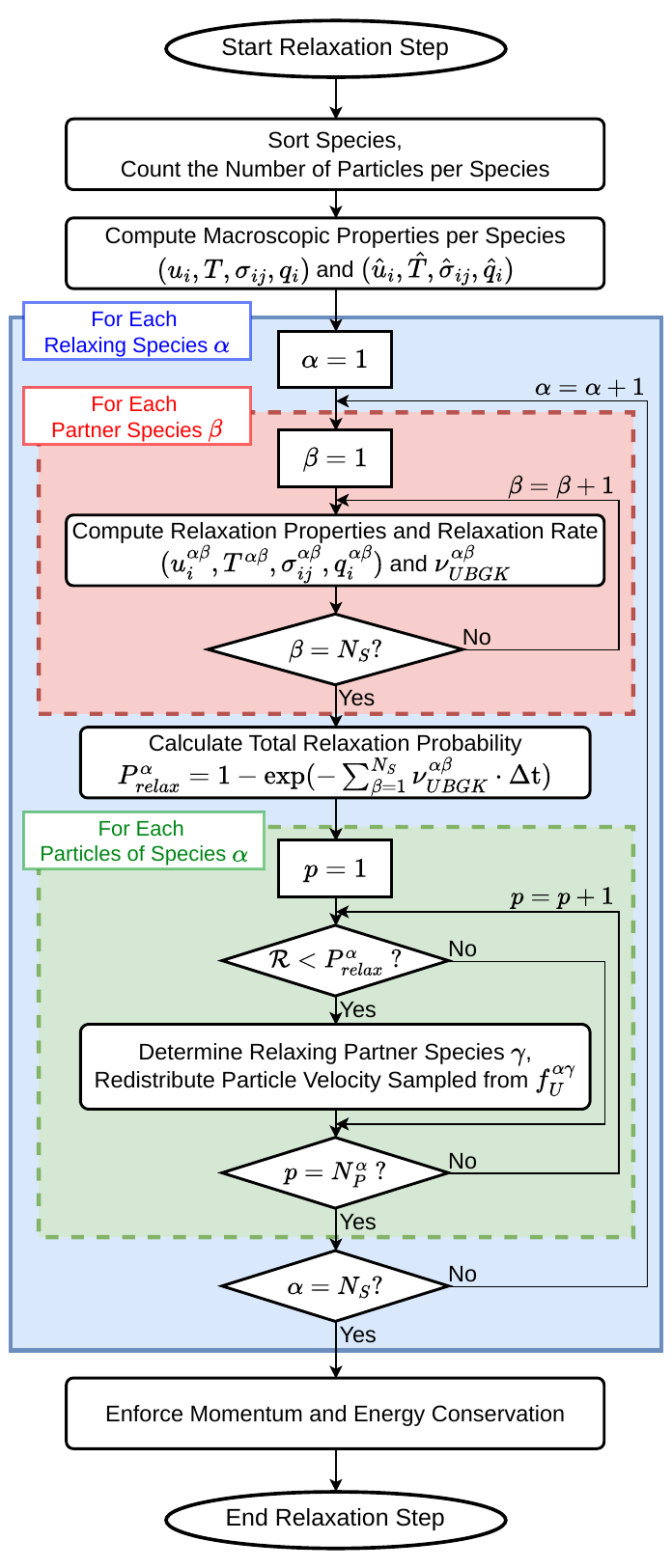}
    \caption{\label{fig:flowchart} Flowchart of the relaxation step for the mixture UBGK model.}
\end{figure*}

For particle simulation, the UBGK model is implemented within SPARTA, an open-source DSMC solver developed by Sandia National Laboratories.\cite{Plimpton2019-SPARTA} A full numerical sequence for the UBGK relaxation step is provided in the flowchart Fig. \ref{fig:flowchart}.

\newpage
\section{\label{sec:results}Results and discussion}


\begin{table}[b]
\caption{\label{table:VHSparameter}VHS molecular parameters.\cite{Bird1994-DSMC}}
\begin{ruledtabular}
\begin{tabular}{l ccccc}
    Species $(\alpha)$ & $m^\alpha \, [\mathrm{kg}]$ & $d_{ref}^\alpha \,[\textrm{m}]$ & $\omega_{ref}^\alpha \,[\textrm{\,-\,}]$ & $T_{ref}^\alpha \,[\textrm{K}]$ \\
    \hline
    He & $ 6.646 \times 10^{-27}$ & $2.331 \times 10^{-10}$ & $0.66$ & $273.15$ \\
    Ne & $ 3.351 \times 10^{-26}$ & $2.766 \times 10^{-10}$ & $0.66$ & $273.15$ \\
    Ar & $ 6.634 \times 10^{-26}$ & $4.170 \times 10^{-10}$ & $0.81$ & $273.15$ \\
    Kr & $ 1.392 \times 10^{-25}$ & $4.762 \times 10^{-10}$ & $0.80$ & $273.15$ \\
\end{tabular}
\end{ruledtabular}
\end{table}

The mixture UBGK model is validated against DSMC through four numerical test cases: homogeneous relaxation, Poiseuille flow, Couette flow, and hypersonic flow around a cylinder. In addition, computational efficiency is assessed by comparing the performance of the UBGK model and DSMC. For all cases, simulated gases are restricted to monatomic species. VHS molecular parameters for the species considered in this study are listed in Table \ref{table:VHSparameter}, where $m^\alpha$ denotes the molecular mass, $d_{ref}^\alpha$ is the reference diameter, $\omega_{ref}^\alpha$ is the viscosity–temperature exponent, and $T_{ref}^\alpha$ is the reference temperature at which $d_{ref}^\alpha$ and $\omega_{ref}^\alpha$ are defined. The intermolecular parameters are evaluated using a collision-averaged approach as:

\begin{equation}
\begin{aligned}
    d_{ref}^{\alpha\beta} = \left( d_{ref}^\alpha + d_{ref}^\beta \right) / 2,
\end{aligned}
\end{equation}
\begin{equation}
\begin{aligned}
    \omega_{ref}^{\alpha\beta} = \left( \omega_{ref}^\alpha + \omega_{ref}^\beta \right) / 2,
\end{aligned}
\end{equation}
\begin{equation}
\begin{aligned}
    T_{ref}^{\alpha\beta} = \left( T_{ref}^\alpha + T_{ref}^\beta \right) / 2.
\end{aligned}
\end{equation}
For a gas mixture, the molecular mean free path $\lambda_{mix}$ and the molecular mean collision time $\tau_{mix}$ are evaluated with\cite{Bird1994-DSMC}
\begin{equation}
\begin{aligned}
    \lambda_{mix}
    &= \sum_{\alpha = 1}^{N_S} \frac{n^\alpha}{n} \Bigg[
        \sum_{\beta = 1}^{N_S} \pi n^\beta \left( d_{ref}^{\alpha\beta} \right)^2 \left( \frac{T_{ref}^{\alpha\beta}}{T} \right)^{\omega^{\alpha\beta} - 0.5} \sqrt{1 + \frac{m^\alpha}{m^\beta}} 
    \Bigg]^{-1},
\end{aligned}
\end{equation}
\begin{equation}
\begin{aligned}
    \tau^{-1}_{mix}
    &= \sum_{\alpha = 1}^{N_S} \frac{n^\alpha}{n} \Bigg[
        \sum_{\beta = 1}^{N_S} 2 \sqrt{\pi} n^\beta \left( d_{ref}^{\alpha\beta} \right)^2 \left( \frac{T}{T_{ref}^{\alpha\beta}} \right)^{1 - \omega^{\alpha\beta}} \sqrt{\frac{2 k_B T_{ref}^{\alpha\beta}}{m_r^{\alpha\beta}}}
    \Bigg].
\end{aligned}
\end{equation}
To ensure that observed differences are solely due to the collision and relaxation models, identical spatial and temporal discretization is employed for both the UBGK model and DSMC. For quantitative evaluation, the relative $\mathrm{L}2$-norm error with respect to DSMC reference solution is obtained as:
\begin{equation}\label{eqn:RelativeL2NormError}
\begin{aligned}
    E_{L2} \left( Y \right) = \frac{\lVert Y_{UBGK} - Y_{DSMC} \rVert_2}{\lVert Y_{DSMC} \rVert_2},
\end{aligned}
\end{equation}
where $Y$ denotes a macroscopic quantity of interest, such as velocity, temperature, shear stress, or heat flux.

\subsection{\label{subsec:HomogeneousRelaxation} Homogeneous relaxation}


A homogeneous relaxation problem is investigated. In the absence of spatial gradients and particle-boundary interactions, the temporal evolution is governed solely by particle-particle interactions, making this case suitable for examining relaxation characteristics. To initialize a non-equilibrium initial state, particle velocities are sampled from Grad's 13-moment distribution function $f_{G13}^\alpha$, given by:
\begin{equation}
\begin{aligned}
    f_{G13}^\alpha = M^\alpha(0) \Bigg( 
        1 
        + \frac{\sigma_{ij}^\alpha(0) C_{\langle i}^\alpha C_{j \rangle}^\alpha}{2 \rho^\alpha \left( R^\alpha T^\alpha(0) \right)^2} 
        + \frac{q_j^\alpha(0)C_j^\alpha}{5 \rho^\alpha \left( R^\alpha T^\alpha(0) \right)^2} \left( \frac{C_j^\alpha C_j^\alpha}{R^\alpha T^\alpha(0)} - 5 \right)
    \Bigg),
\end{aligned}\end{equation}
where $C_i^\alpha = c_i^\alpha - u_i^\alpha(0)$ is the thermal velocity, and $u_i^\alpha(0)$, $T^\alpha(0)$, $\sigma_{ij}^\alpha(0)$, and $q_i^\alpha(0)$ represent the initial conditions for velocity, temperature, shear stress and heat flux, respectively. Throughout this problem, the gas mixture consists of Neon and Argon with equal number densities, $n^{Ne} = n^{Ar} = 5 \times 10^{19} \, \textrm{m}^{-3}$. The initial conditions for Neon and Argon are listed in Table \ref{table:HomRel_IC}. The simulation employs a single computational cell of size $\Delta x = 0.01~\mathrm{m}$, with a total of $1 \times 10^{8}$ simulated particles. The time step is set to $\Delta t = 0.01~\tau_{mix} = 4.0 \times 10^{-7}~\mathrm{s}$. Relative $\mathrm{L}2$-norm errors are computed over the entire time history for quantitative assessment.

\begin{table}[h]
\caption{\label{table:HomRel_IC}Initial conditions for homogeneous relaxation problems.}
\begin{ruledtabular}
    \begin{tabular}{cccccc}
        Case & Species $(\alpha)$ & $u_x^\alpha(0)\,[\mathrm{m/s}]$ & $T^\alpha(0)\,[\mathrm{K}]$ & $\sigma_{ij}^\alpha(0)$ & $q_i^\alpha(0)$ \\
        \hline
        \multirow{2}{*}{A}
        & Ne & $100$ & $200$ & $0.1\,\rho^{Ne} R^{Ne} T^{Ne}$ & $0.1\,\rho^{Ne} \left( R^{Ne} T^{Ne} \right)^{1.5}$ \\
        & Ar & $0$   & $300$ & $0.1\,\rho^{Ar} R^{Ar} T^{Ar}$ & $0.1\,\rho^{Ar} \left( R^{Ar} T^{Ar} \right)^{1.5}$ \\
    \end{tabular}
\end{ruledtabular}
\end{table}

\subsubsection{\label{subsubsec:HomogeneousRelaxation-subcase1} Sensitivity analysis with respect to $C_{ES}$ and $C_S$}


\begin{figure}[t]
    \centering
	\begin{subfigure}{0.495\textwidth}
		\centering
        \includegraphics[width=1.0\linewidth]{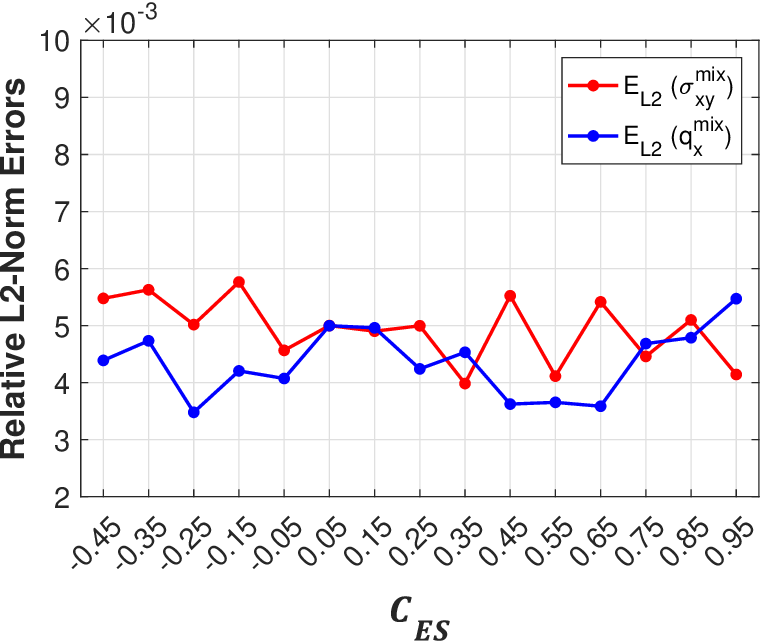}
	\end{subfigure}
	\caption{\label{fig:1-1_Relax}Relative L2-norm errors of mixture shear stress and heat flux as a function of the parameter $C_{ES}$ in homogeneous relaxation.}
\end{figure}

The parameters $C_{ES}$ and $C_S$, introduced in Eqs. (\ref{eqn:UBGK-TargetDistribution})$-$(\ref{eqn:UBGK-TargetDistribution_S}), control the relative contributions of shear stress and heat flux corrections in the UBGK model. Since these parameters are not uniquely determined, their values may influence non-equilibrium flow predictions, making sensitivity analysis essential.\cite{Chen2015-UBGK} Accordingly, within the admissible range ensuring positive definiteness of $\lambda'^{\alpha\beta}$, $C_{ES}$ is varied from $-0.45$ to $0.95$ with an increment of $0.1$, and $C_S$ is set to enforce the target Prandtl number of $2/3$. Figure \ref{fig:1-1_Relax} presents the relative $\mathrm{L}2$-norm errors of $\sigma_{xy}^{mix}$ and $q_x^{mix}$ as a function of $C_{ES}$. Across the examined range, both quantities exhibit consistently low levels of error, remaining below $0.006$, indicating that the model's performance is insensitive to the choice of $C_{ES}$. Although minor oscillations are observed, they are attributed to numerical artifacts rather than physical sensitivity. A representative value of $C_{ES} = -0.25$ is selected for subsequent simulations.

\subsubsection{\label{subsubsec:HomogeneousRelaxation_subcase2} Accuracy evaluation of temporal evolution}


\begin{figure*}[t]
	\centering
	\begin{subfigure}{0.485\textwidth}
		\centering
		\includegraphics[width=1.\linewidth]{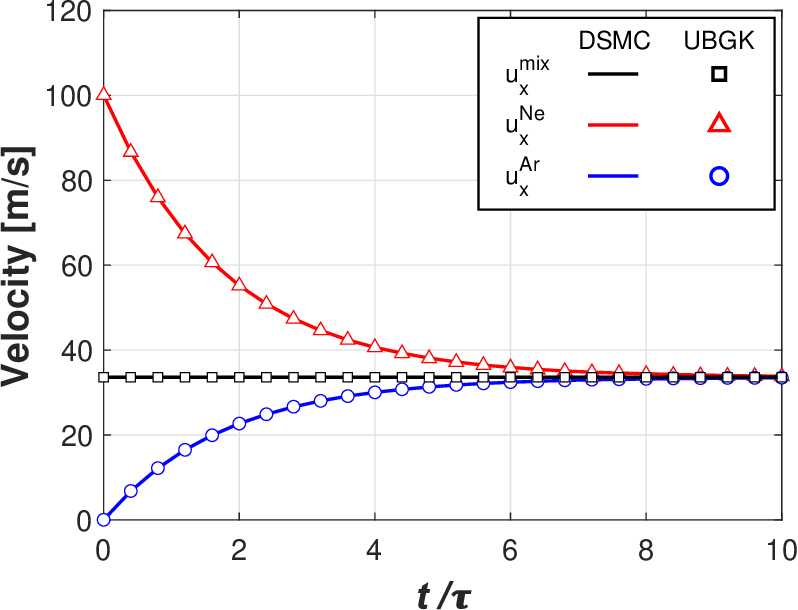}
            \caption{\label{subfig:1-2_Relax-Velo}Velocity.}
	\end{subfigure}
	\begin{subfigure}{0.485\textwidth}
		\centering
		\includegraphics[width=1.\linewidth]{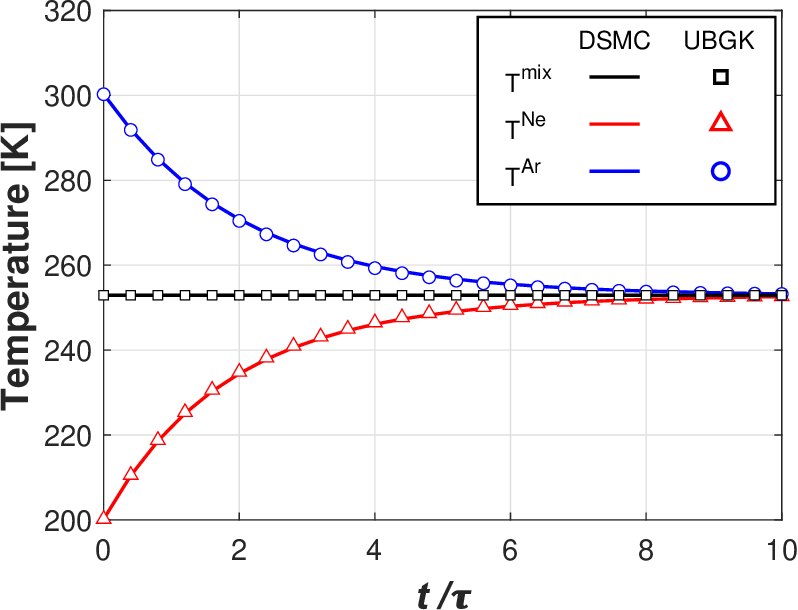}
            \caption{\label{subfig:1-2_Relax-Temp}Temperature.}
	\end{subfigure}
	\centering
	\begin{subfigure}{0.495\textwidth}
		\centering
		\includegraphics[width=1.\linewidth]{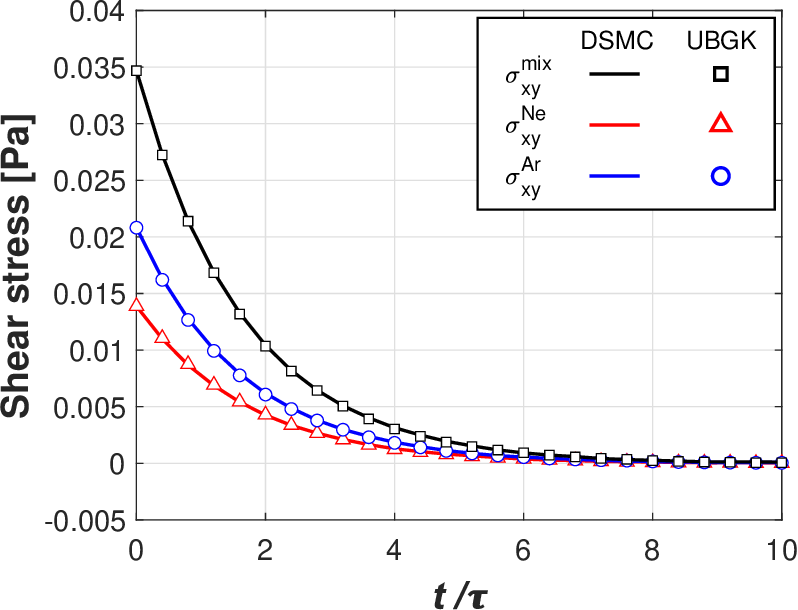}
            \caption{\label{subfig:1-2_Relax-Shear}Shear stress.}
	\end{subfigure}
	\begin{subfigure}{0.485\textwidth}
		\centering
		\includegraphics[width=1.\linewidth]{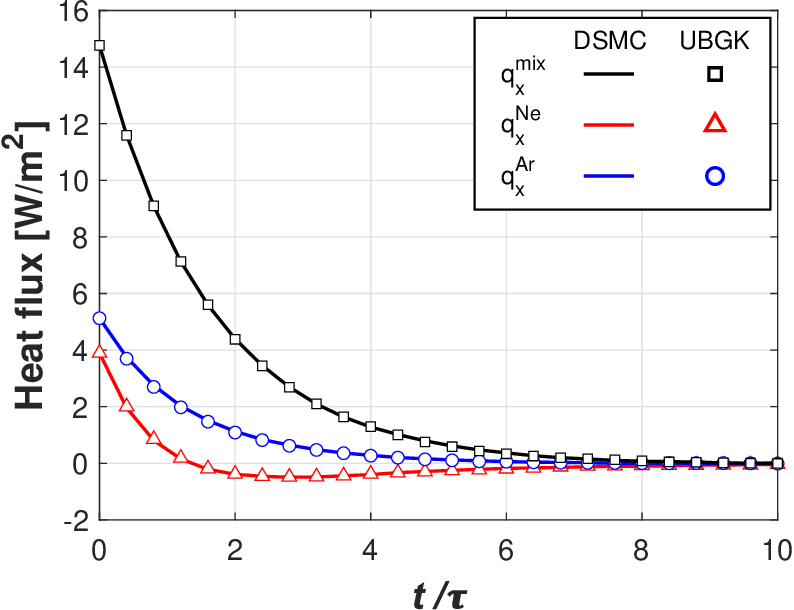}
            \caption{\label{subfig:1-2_Relax-heat}Heat flux.}
	\end{subfigure}
	\caption{\label{fig:1-2_Relax}Temporal evolution of velocity, temperature, shear stress, and heat flux for the mixture, Neon, and Argon of homogeneous relaxation.}
\end{figure*}

The accuracy of the homogeneous relaxation solution is evaluated by comparing the UBGK prediction against DSMC. The comparison focuses on $u_x^\alpha$, $T^\alpha$, $\sigma_{xy}^\alpha$, and $q_x^\alpha$ of the mixture as well as individual species. Figure \ref{fig:1-2_Relax} illustrates the relaxation process of the macroscopic quantities obtained from the UBGK model and DSMC. Figures \ref{subfig:1-2_Relax-Velo} and \ref{subfig:1-2_Relax-Temp} show the relaxation of $u_x^\alpha$ and $T^\alpha$, respectively. Since the system's total momentum and energy are conserved, the velocities and temperatures of individual species relax toward the corresponding mixture properties. The relaxations of $\sigma_{xy}^\alpha$ and $q_x^\alpha$ are illustrated in Figs. \ref{subfig:1-2_Relax-Shear} and \ref{subfig:1-2_Relax-heat}, where both quantities decay toward zero due to dissipation, as the system relaxes toward thermodynamic equilibrium. Across all quantities considered, no phase lag in the relaxation process is observed, and the UBGK model predictions closely follow DSMC results. The resulting $\textrm{L}2$-norm error levels shown in Table \ref{table:1-2_Relax-RelL2Error} are found to be consistently small, remaining below $0.03$ for all components, confirming that the UBGK model accurately predicts the temporal evolution of macroscopic properties.

\begin{table}[t]
\caption{\label{table:1-2_Relax-RelL2Error}Relative L2-norm errors of velocity, temperature, shear stress, and heat flux in homogeneous relaxation.}
\begin{ruledtabular}
    \begin{tabular}{l cccc}
            \multirow{2}{*}{Component} &
            \multicolumn{4}{c}{Relative L2-norm errors} \\
            \cline{2-5} 
             &
            $E_{L2} \left( u_x^\alpha \right)$ &
            $E_{L2} \left( T^\alpha \right)$ &
            $E_{L2} \left( \sigma_{xy}^\alpha \right)$ &
            $E_{L2} \left( q_x^\alpha \right)$ \\
        \hline
            Ne       & 0.00125 & 0.00111 & 0.00831 & 0.02200 \\
            Ar       & 0.00109 & 0.00102 & 0.00885 & 0.01590 \\
            Mixture  & -       & -       & 0.00502 & 0.00348 \\
    \end{tabular}
\end{ruledtabular}
\end{table}

\subsection{\label{subsec:Poiseuille}Poiseuille flow}


A 1D Poiseuille flow is considered as a canonical internal flow problem to assess the model performance for near-equilibrium flows. The flow is driven by a constant streamwise pressure-density acceleration of $\left| \nabla p / \rho \right| = 6180~\mathrm{m/s^2}$. The two parallel, horizontal plates are separated by a distance $L = 1~\mathrm{m}$, and are maintained at a uniform temperature of $300~\mathrm{K}$. Both walls are stationary and are assumed to be fully diffusive. Simulations are conducted at a fixed Knudsen number of $\mathrm{Kn} = 0.01$. The working gases are $(\mathrm{Ne-Ar})$ mixtures with three different mole fractions, $\left( \chi^{\mathrm{Ne}}, \chi^{\mathrm{Ar}} \right) \in \left\{ (0.25, 0.75),\ (0.50, 0.50),\ (0.75, 0.25) \right\}$. The computational domain is discretized into $1 \times 1000$ cells, which corresponds to $\Delta y = 0.1~\lambda_{mix}$, and a time step of $\Delta t = 0.05~\tau_{mix} = 1 \times 10^{-6}~\mathrm{s}$ is implemented. In order to reduce statistical noise, a total of $1 \times 10^{7}$ particles are employed. Detailed flow conditions are summarized in Table \ref{table:2_Poiseuille-Condition}.

\begin{table}[b]
\caption{\label{table:2_Poiseuille-Condition}Flow conditions for Poiseuille flow.}
\begin{ruledtabular}
\begin{tabular}{ccccc}
    Case & $\chi^{\mathrm{Ne}} \, [\textrm{\,-\,}]$ & $\chi^{\mathrm{Ar}} \, [\textrm{\,-\,}]$ & $n^{\mathrm{Ne}} \, [\mathrm{m}^{-3}]$ & $n^{\mathrm{Ar}} \, [\mathrm{m}^{-3}]$ \\
    \hline
    B-1 & $0.25$ & $0.75$ & $ 4.070 \times 10^{19}$ & $1.221 \times 10^{20}$ \\
    B-2 & $0.50$ & $0.50$ & $ 9.940 \times 10^{19}$ & $9.940 \times 10^{19}$ \\
    B-3 & $0.75$ & $0.25$ & $ 1.824 \times 10^{20}$ & $6.080 \times 10^{19}$ \\
\end{tabular}
\end{ruledtabular}
\end{table}


\begin{figure*}[t]
    \centering
	\begin{subfigure}{0.32\textwidth}
		\centering
		\includegraphics[width=1.\linewidth]{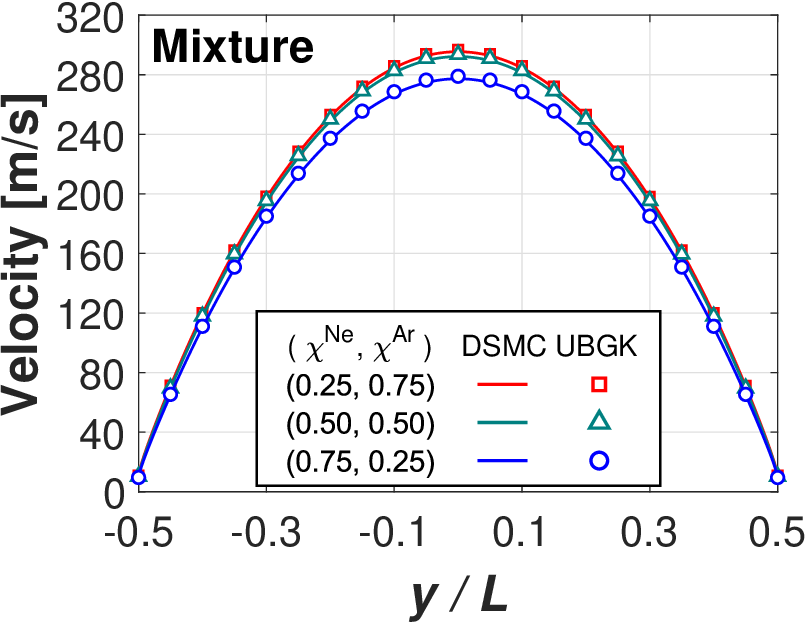}
            \caption{\label{subfig:2_Poiseuille-MixVelo}Mixture velocity.}
	\end{subfigure}
	\begin{subfigure}{0.32\textwidth}
		\centering
		\includegraphics[width=1.\linewidth]{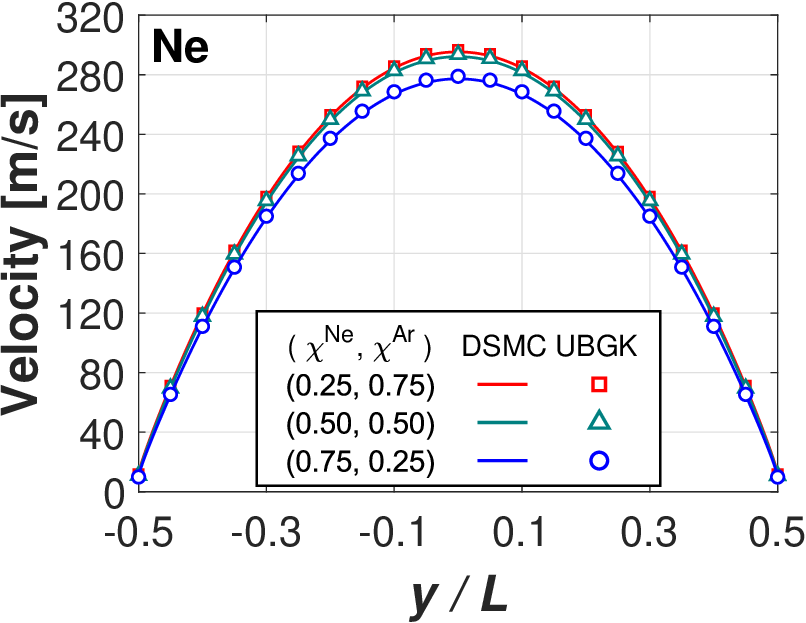}
            \caption{\label{subfig:2_Poiseuille-NeVelo}Neon velocity.}
	\end{subfigure}
        \begin{subfigure}{0.32\textwidth}
		\centering
		\includegraphics[width=1.\linewidth]{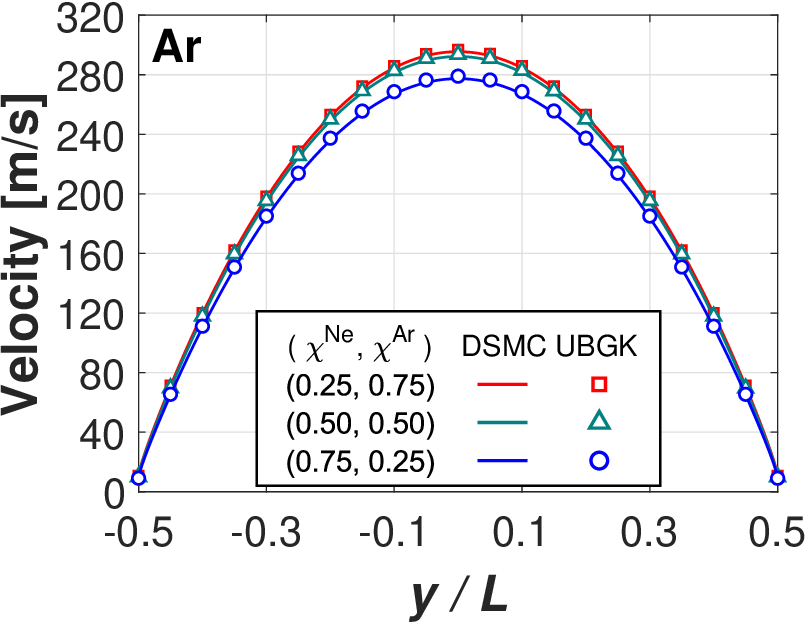}
            \caption{\label{subfig:2_Poiseuille-ArVelo}Argon velocity.}
	\end{subfigure}
    \centering
	\begin{subfigure}{0.32\textwidth}
		\centering
		\includegraphics[width=1.\linewidth]{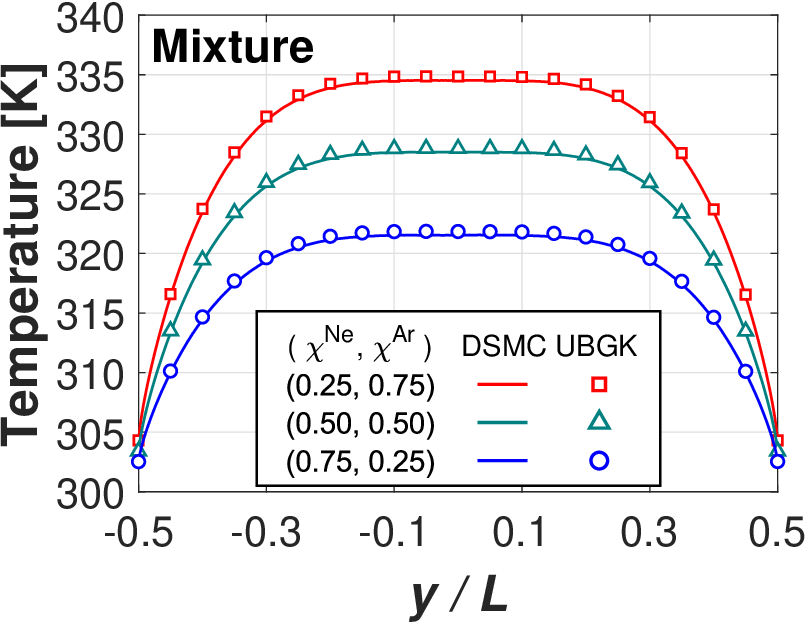}
            \caption{\label{subfig:2_Poiseuille-MixTemp}Mixture temperature.}
	\end{subfigure}
	\begin{subfigure}{0.32\textwidth}
		\centering
		\includegraphics[width=1.\linewidth]{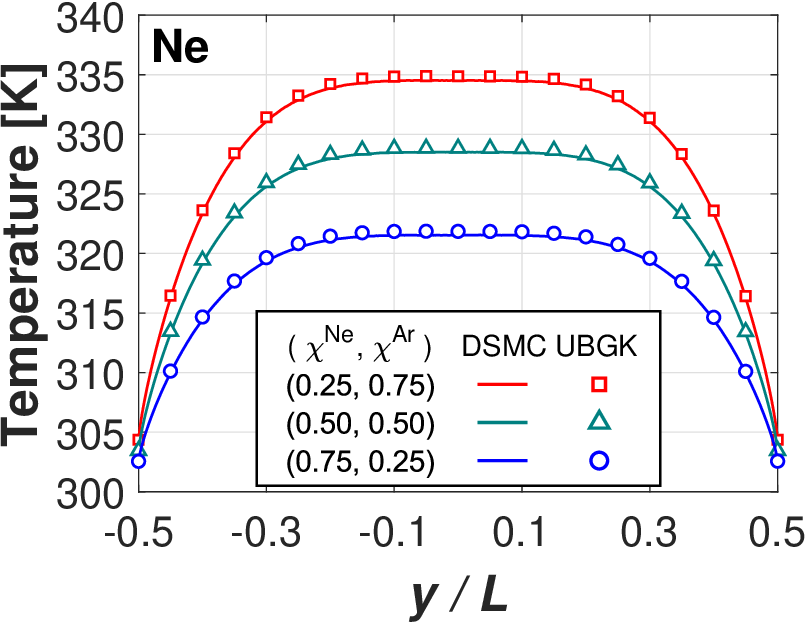}
            \caption{\label{subfig:2_Poiseuille-NeTemp}Neon temperature.}
	\end{subfigure}
        \begin{subfigure}{0.32\textwidth}
		\centering
		\includegraphics[width=1.\linewidth]{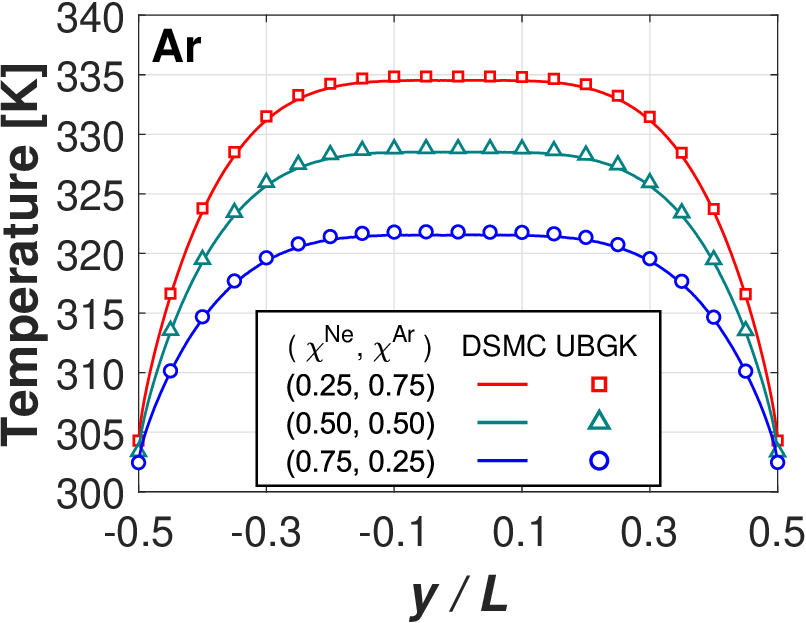}
            \caption{\label{subfig:2_Poiseuille-ArTemp}Argon temperature.}
	\end{subfigure}
    \centering
	\begin{subfigure}{0.32\textwidth}
		\centering
		\includegraphics[width=1.\linewidth]{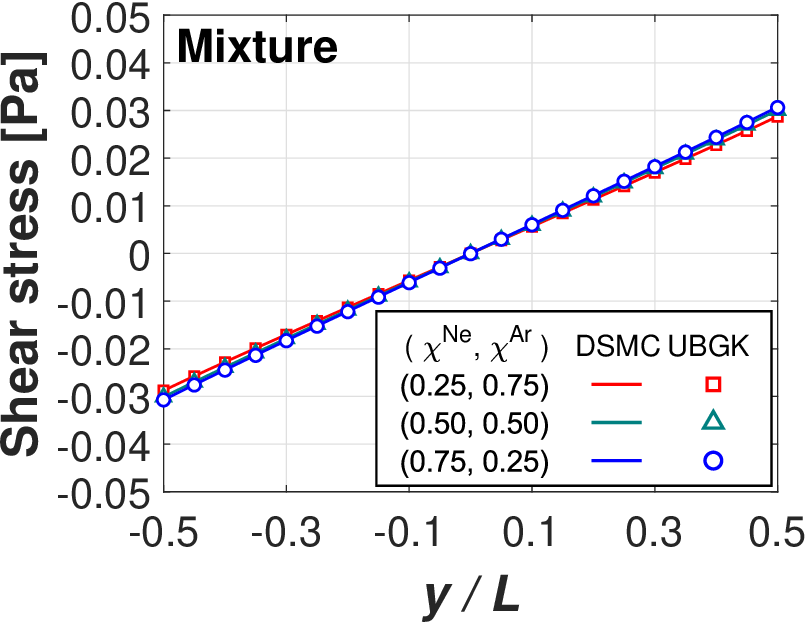}
            \caption{\label{subfig:2_Poiseuille-MixShear}Mixture shear stress.}
	\end{subfigure}
	\begin{subfigure}{0.32\textwidth}
		\centering
		\includegraphics[width=1.\linewidth]{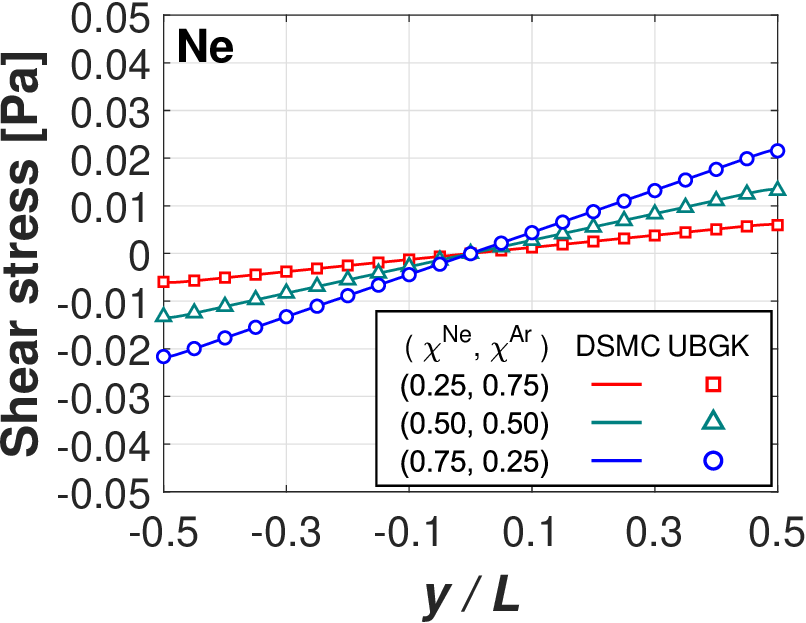}
            \caption{\label{subfig:2_Poiseuille-NeShear}Neon shear stress.}
	\end{subfigure}
        \begin{subfigure}{0.32\textwidth}
		\centering
		\includegraphics[width=1.\linewidth]{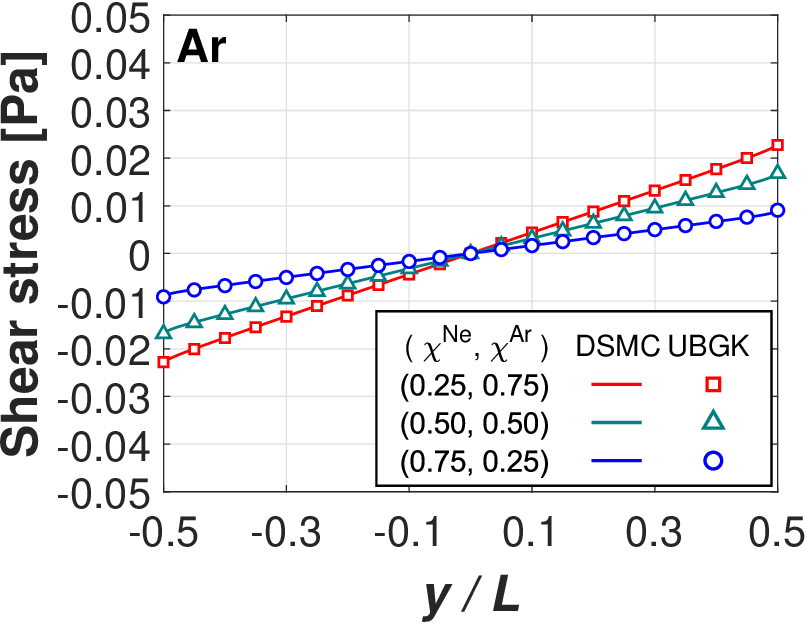}
            \caption{\label{subfig:2_Poiseuille-ArShear}Argon shear stress.}
	\end{subfigure}
    \caption{\label{fig:2_Poiseuille}Profiles of velocity, temperature, and shear stress for the mixture, Neon, and Argon of Poiseuille flow.}
\end{figure*}

Figure \ref{fig:2_Poiseuille} compares $u_x^\alpha$, $T^\alpha$, and $\sigma_{xy}^\alpha$ profiles for the mixture and individual species, and Table \ref{table:2_Poiseuille-RelL2Error} summarizes the corresponding relative $\mathrm{L}2$-norm errors over the entire computational domain. Figures \ref{subfig:2_Poiseuille-MixVelo}$-$\ref{subfig:2_Poiseuille-ArVelo} illustrate $u_x^\alpha$ profiles, and exhibit the characteristic parabolic distribution, as well as small but finite velocity slip near the walls. The profiles of the mixture and individual species are indistinguishable from one another, as non-equilibrium effects remain weak under near-continuum conditions. Figures \ref{subfig:2_Poiseuille-MixTemp}$-$\ref{subfig:2_Poiseuille-ArTemp} illustrate $T^\alpha$ profiles, and show the expected quartic profile, along with temperature jumps near the walls. As with the $u_x^\alpha$ profiles, $T^\alpha$ profiles of the mixture and individual species are indistinguishable from one another under near-continuum conditions. Figures \ref{subfig:2_Poiseuille-MixShear}$-$\ref{subfig:2_Poiseuille-ArShear} illustrate $\sigma_{xy}^\alpha$ profiles, and display the expected linear variation across the channel. Unlike $u_x^\alpha$ and $T^\alpha$ profiles, $\sigma_{xy}^\alpha$ profiles exhibit noticeable differences between Neon and Argon, as well as the mixture. This difference arises from the mass-weighted second-order definition of $\sigma_{xy}^\alpha$, i.e. Eq. (\ref{eqn:shearstress-species}), and the mole fraction of each species. While $u_x^\alpha$ and $T^\alpha$ are defined as averaged quantities, $\sigma_{xy}^\alpha$ is obtained as a summation of second-order moments. As a result, species with different molecular masses and mole fractions exhibit distinct contributions to $\sigma_{xy}^\alpha$, leading to the observed discrepancies between Neon, Argon, and the mixture. Across all cases and quantities considered, the UBGK model closely follows the DSMC results, with the relative $\mathrm{L}2$-norm errors remaining below $0.008$. These results demonstrate that the UBGK model recovers accurate transport characteristics in the near-continuum regime over a range of mixture compositions.

\begin{table}[t]
\caption{\label{table:2_Poiseuille-RelL2Error}Relative L2-norm errors of velocity, temperature, and shear stress in Poiseuille flow.}
\begin{ruledtabular}
\begin{tabular}{l l ccc}
    \multirow{2}{*}{Case} &
    \multirow{2}{*}{Component $(\alpha)$} &
    \multicolumn{3}{c}{Relative L2-norm errors} \\
    \cline{3-5}
     &  &
    $E_{L2} \left( u_x^\alpha \right)$ &
    $E_{L2} \left( T^\alpha \right)$ &
    $E_{L2} \left( \sigma_{xy}^\alpha \right)$ \\
    \hline
    \multirow{3}{*}{B-1 $(\chi^{\mathrm{Ne}}=0.25)$} 
        & Ne      & 0.0043 & 0.0009 & 0.0022 \\
        & Ar      & 0.0043 & 0.0009 & 0.0079 \\
        & Mixture & 0.0043 & 0.0009 & 0.0030 \\
    \hline
    \multirow{3}{*}{B-2 $(\chi^{\mathrm{Ne}}=0.50)$} 
        & Ne      & 0.0052 & 0.0009 & 0.0025 \\
        & Ar      & 0.0052 & 0.0009 & 0.0059 \\
        & Mixture & 0.0051 & 0.0008 & 0.0048 \\
    \hline
    \multirow{3}{*}{B-3 $(\chi^{\mathrm{Ne}}=0.75)$} 
        & Ne      & 0.0063 & 0.0008 & 0.0030 \\
        & Ar      & 0.0064 & 0.0009 & 0.0046 \\
        & Mixture & 0.0063 & 0.0007 & 0.0079 \\
\end{tabular}
\end{ruledtabular}
\end{table}

\subsection{\label{subsec:Couette}Couette flow}


\begin{table}[b]
\caption{\label{table:3_Couette-Condition}Flow and numerical conditions for Couette flow.}
\begin{ruledtabular}
\begin{tabular}{ccccc}
    Case & Kn & $n^{\mathrm{Ne}}, n^{\mathrm{Ar}} \, [\mathrm{m}^{-3}]$ & Cell size $\, [\mathrm{m}]$ & Time step $\, [\mathrm{s}]$ \\
    \hline
    C-1 & $0.005$ & $1.989 \times 10^{20}$ & $1~\times~10^{-3}$ & $1~\times~10^{-6}$ \\
    C-2 & $0.05$  & $1.989 \times 10^{19}$ & $1~\times~10^{-2}$ & $1~\times~10^{-5}$ \\
    C-3 & $0.5$   & $1.989 \times 10^{18}$ & $1~\times~10^{-2}$ & $1~\times~10^{-5}$ \\
    C-4 & $1.5$   & $6.630 \times 10^{17}$ & $1~\times~10^{-2}$ & $1~\times~10^{-5}$ \\
\end{tabular}
\end{ruledtabular}
\end{table}

A 1D Couette flow is investigated to assess the UBGK model as $\mathrm{Kn}$ increases from near-equilibrium toward more rarefied conditions. The flow is confined between two parallel, horizontal plates, which are separated by a distance $L = 1~\mathrm{m}$. The flow is driven by the relative motion of the walls, with the upper plate moving at $300~\mathrm{m/s}$ and the lower plate moving at $-300~\mathrm{m/s}$. Both walls are maintained at a uniform temperature of $300~\mathrm{K}$, and are assumed to be fully diffusive. Simulations are conducted for four different Knudsen numbers, $\mathrm{Kn} \in \{ 0.005, 0.05, 0.5, 1.5 \}$. The working gas is a $(\mathrm{Ne-Ar})$ mixture at a fixed mole fraction of $\left( \chi^{\mathrm{Ne}}, \chi^{\mathrm{Ar}} \right) = (0.50, 0.50)$. The computational domain is discretized such that the cell size satisfies $\Delta y < 0.2~\lambda_{mix}$, and the time step is chosen to satisfy $\Delta t < 0.2~\tau_{mix}$. $10000$ particles per cell are generated to reduce statistical noise. Detailed flow and numerical conditions are summarized in Table \ref{table:3_Couette-Condition}.


\begin{figure*}[t]
    \centering
	\begin{subfigure}{0.32\textwidth}
		\centering
		\includegraphics[width=1.\linewidth]{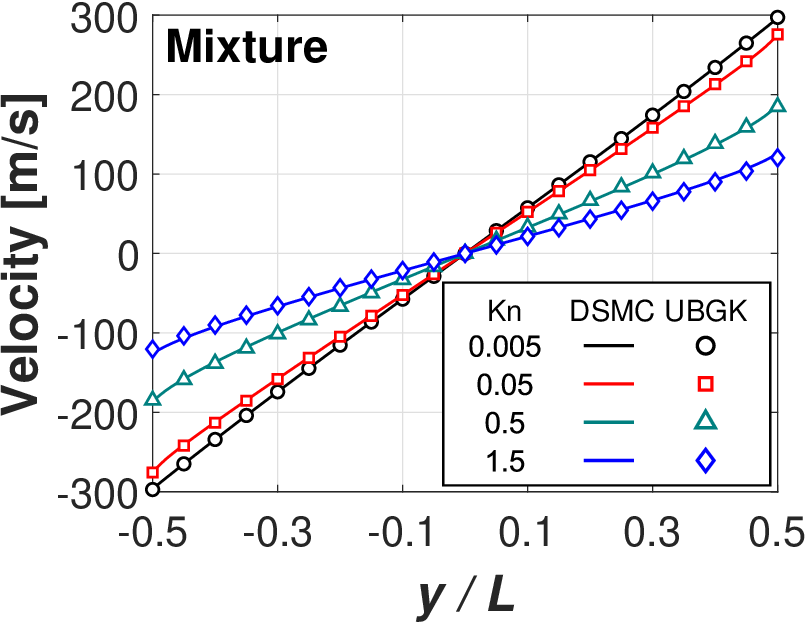}
            \caption{\label{subfig:3_Couette-MixVelo}Mixture velocity.}
	\end{subfigure}
	\begin{subfigure}{0.32\textwidth}
		\centering
		\includegraphics[width=1.\linewidth]{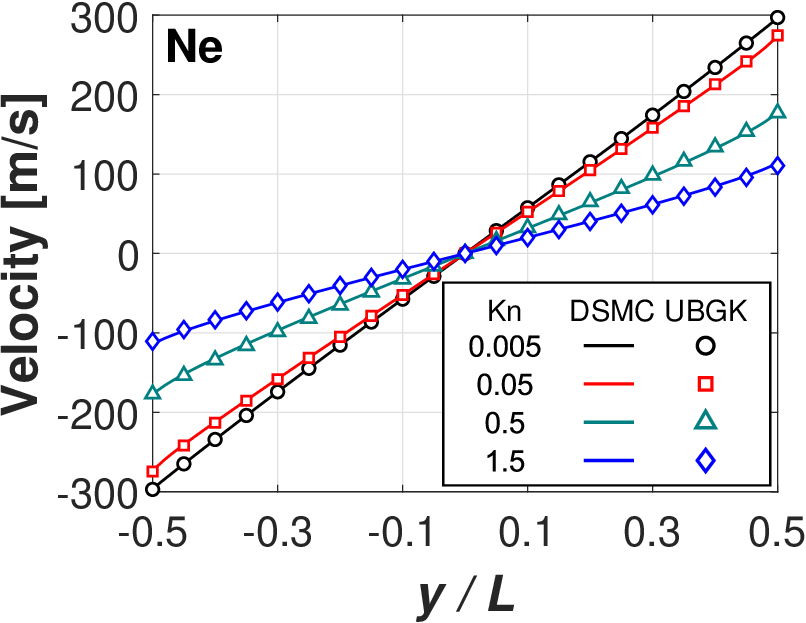}
            \caption{\label{subfig:3_Couette-NeVelo}Neon velocity.}
	\end{subfigure}
        \begin{subfigure}{0.32\textwidth}
		\centering
		\includegraphics[width=1.\linewidth]{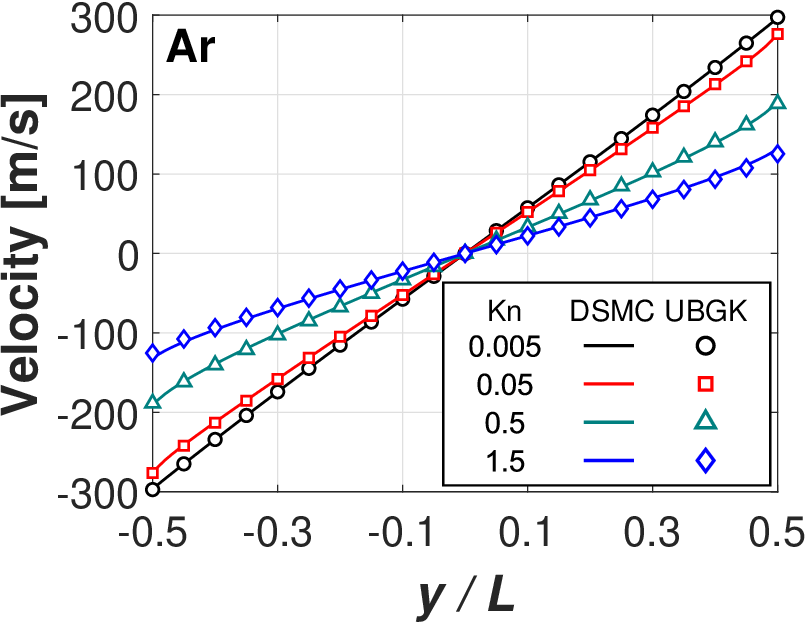}
            \caption{\label{subfig:3_Couette-ArVelo}Argon velocity.}
	\end{subfigure}
    \centering
	\begin{subfigure}{0.32\textwidth}
		\centering
		\includegraphics[width=1.\linewidth]{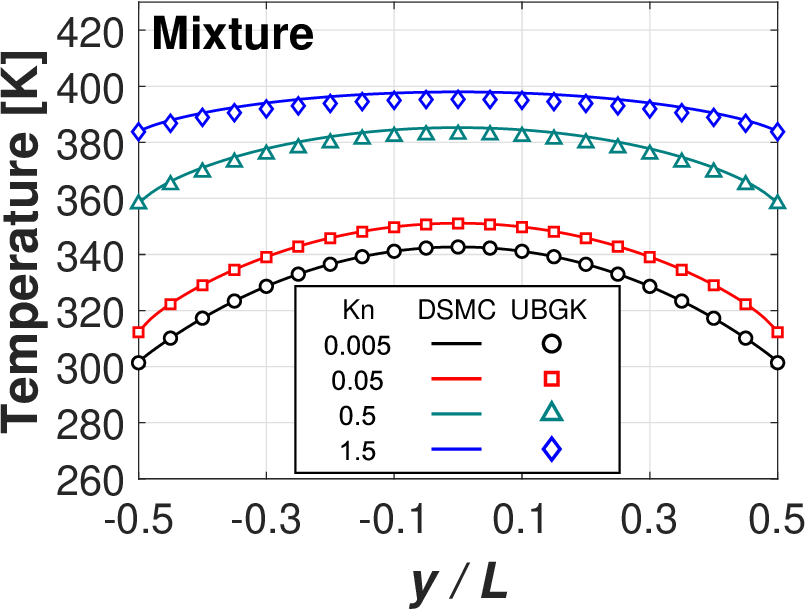}
            \caption{\label{subfig:3_Couette-MixTemp}Mixture temperature.}
	\end{subfigure}
	\begin{subfigure}{0.32\textwidth}
		\centering
		\includegraphics[width=1.\linewidth]{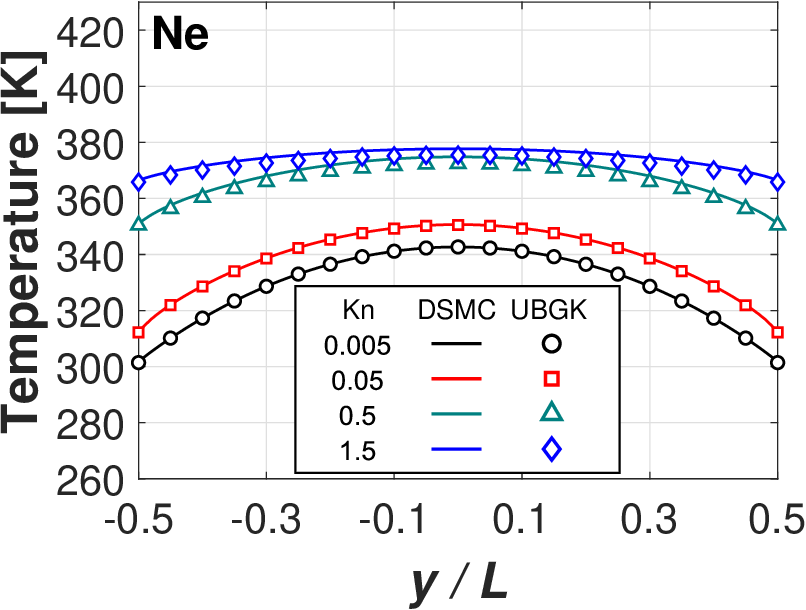}
            \caption{\label{subfig:3_Couette-NeTemp}Neon temperature.}
	\end{subfigure}
        \begin{subfigure}{0.32\textwidth}
        \centering
        \includegraphics[width=1.\linewidth]{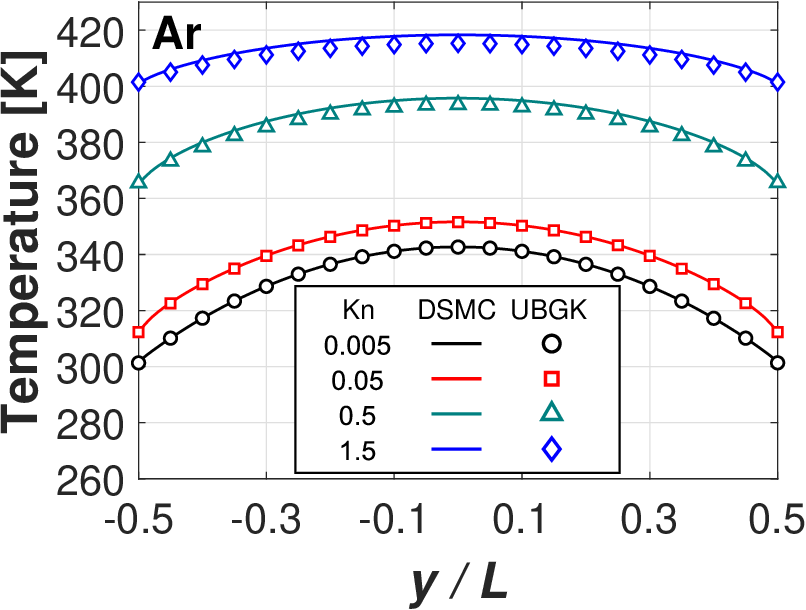}
            \caption{\label{subfig:3_Couette-ArTemp}Argon temperature.}
	\end{subfigure}
    \centering
	\begin{subfigure}{0.32\textwidth}
		\centering
		\includegraphics[width=1.\linewidth]{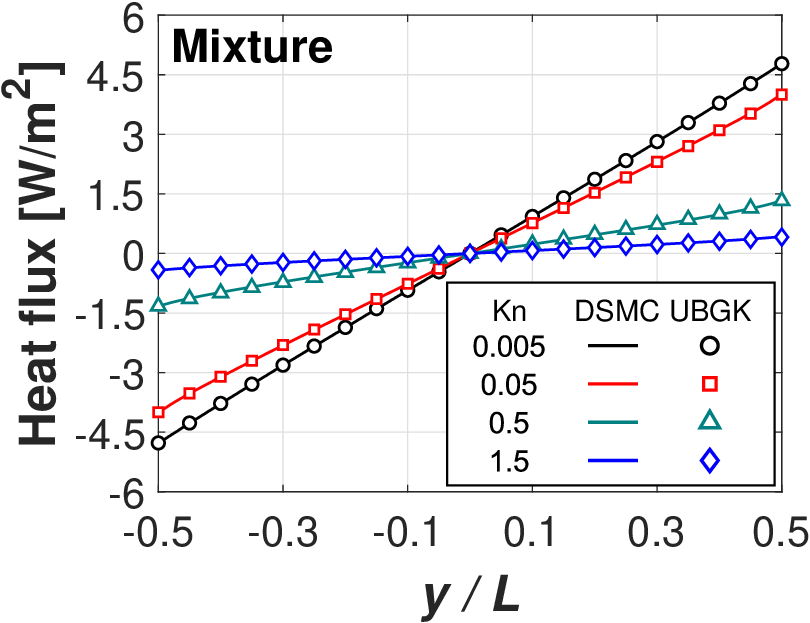}
            \caption{\label{subfig:3_Couette-MixHeat}Mixture heat flux.}
	\end{subfigure}
	\begin{subfigure}{0.32\textwidth}
		\centering
		\includegraphics[width=1.\linewidth]{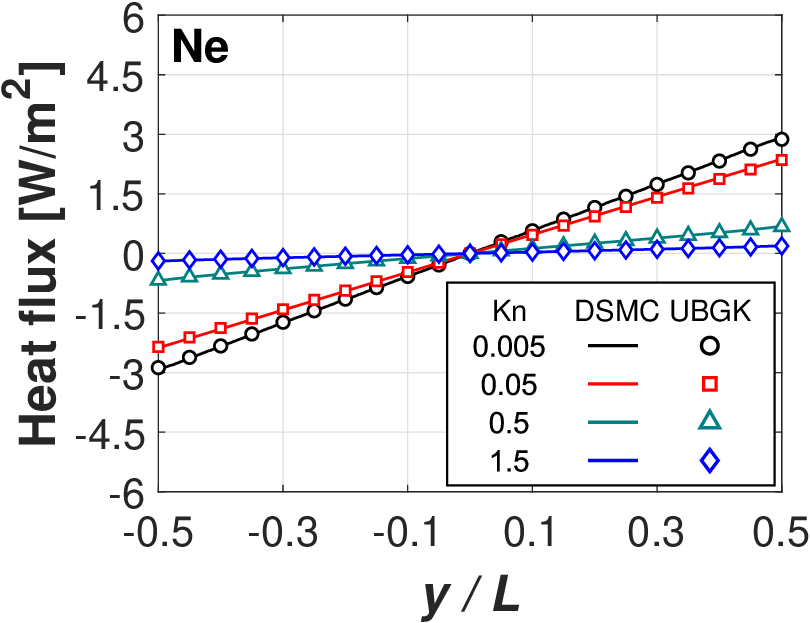}
            \caption{\label{subfig:3_Couette-NeHeat}Neon heat flux.}
	\end{subfigure}
    \begin{subfigure}{0.32\textwidth}
		\centering
		\includegraphics[width=1.\linewidth]{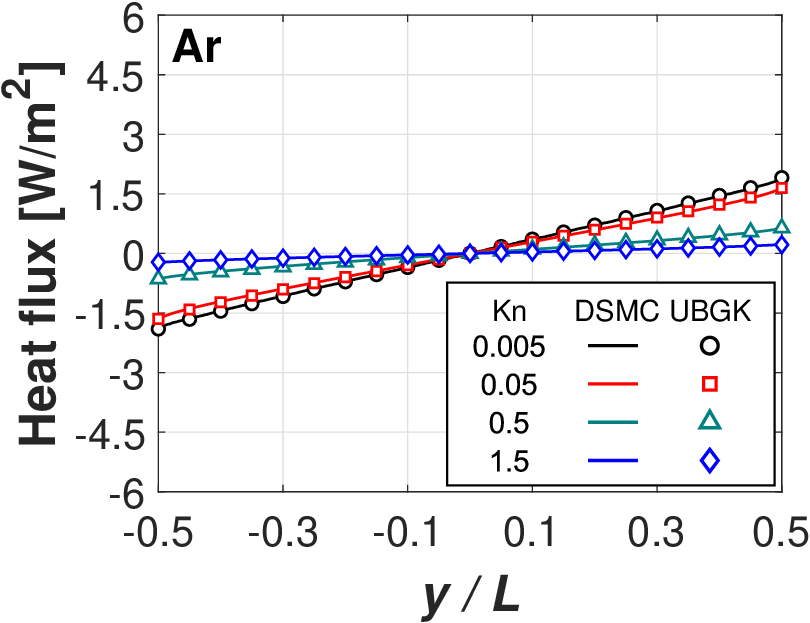}
            \caption{\label{subfig:3_Couette-ArHeat}Argon heat flux.}
	\end{subfigure}
    \caption{\label{fig:3_Couette}Profiles of velocity, temperature, and heat flux for the mixture, Neon, and Argon of Couette flow.}
\end{figure*}

\begin{table}[t]
\caption{\label{table:3_Couette-RelL2Error}Relative L2-norm errors of velocity, temperature, and heat flux in Couette flow.}
\begin{ruledtabular}
\begin{tabular}{l l ccc}
    \multirow{2}{*}{Case} &
    \multirow{2}{*}{Component $(\alpha)$} &
    \multicolumn{3}{c}{Relative L2-norm errors} \\
    \cline{3-5}
     &  &
    $E_{L2} \left( u_x^\alpha \right)$ &
    $E_{L2} \left( T^\alpha \right)$ &
    $E_{L2} \left( q_y^\alpha \right)$ \\
    \hline
    \multirow{3}{*}{C-1 $(\mathrm{Kn}=0.005)$} 
        & Ne      & 0.0003 & 0.0001 & 0.0021 \\
        & Ar      & 0.0003 & 0.0001 & 0.0073 \\
        & Mixture & 0.0004 & 0.0001 & 0.0097 \\
    \hline
    \multirow{3}{*}{C-2 $(\mathrm{Kn}=0.05)$} 
        & Ne      & 0.0036 & 0.0004 & 0.0028 \\
        & Ar      & 0.0040 & 0.0004 & 0.0134 \\
        & Mixture & 0.0034 & 0.0004 & 0.0141 \\
    \hline
    \multirow{3}{*}{C-3 $(\mathrm{Kn}=0.5)$} 
        & Ne      & 0.0114 & 0.0053 & 0.0146 \\
        & Ar      & 0.0145 & 0.0056 & 0.0086 \\
        & Mixture & 0.0101 & 0.0050 & 0.0399 \\
    \hline
    \multirow{3}{*}{C-4 $(\mathrm{Kn}=1.5)$} 
        & Ne      & 0.0253 & 0.0055 & 0.0183 \\
        & Ar      & 0.0167 & 0.0050 & 0.0259 \\
        & Mixture & 0.0292 & 0.0059 & 0.0129 \\
\end{tabular}
\end{ruledtabular}
\end{table}

Figure \ref{fig:3_Couette} compares $u_x^\alpha$, $T^\alpha$, and $q_y^\alpha$ profiles for the mixture and individual species, and Table \ref{table:3_Couette-RelL2Error} summarizes the corresponding relative $\mathrm{L}2$-norm errors over the entire computational domain. Figures \ref{subfig:3_Couette-MixVelo}$-$\ref{subfig:3_Couette-ArVelo} illustrate the $u_x^\alpha$ profiles. At low Knudsen numbers of $\mathrm{Kn}=0.005$ and $0.05$, the profiles remain linear across the channel and are indistinguishable among both species and the mixture. As $\mathrm{Kn}$ increases to $0.5$ and $1.5$, however, non-equilibrium effects become dominant. Nonlinear profiles develop, larger velocity slip near the wall is observed, and the profiles of Neon and Argon, as well as the mixture, become different to one another. Figures \ref{subfig:3_Couette-MixTemp}$-$\ref{subfig:3_Couette-ArTemp} illustrate $T^\alpha$ profiles. At low $\mathrm{Kn}$, the profiles of the mixture and individual species coincide. With increasing rarefaction, however, the mixture and individual species temperatures become distinct, indicating the breakdown of local thermal equilibrium. Figures \ref{subfig:3_Couette-MixHeat}$-$\ref{subfig:3_Couette-ArHeat} illustrate the $q_y^\alpha$ distribution, where differences among Neon, Argon, and the mixture are observed even at low Knudsen numbers. In contrast to $u_x^\alpha$ and $T^\alpha$, $q_y^\alpha$ is defined as a summation of third-order moments, i.e. Eq. (\ref{eqn:heatflux-species}). Consequently, species with different molecular masses exhibit distinct contributions to $q_y^\alpha$, resulting in distinct profiles for Neon, Argon, and the mixture even at low $\mathrm{Kn}$. Across all cases and quantities considered, the UBGK model closely follows DSMC results, reproducing the overall trends and emergence of differences between the mixture and individual species. While the relative $\mathrm{L}2$-norm errors for low $\mathrm{Kn}$ remain below $0.0141$, they increase up to $0.0399$ for high $\mathrm{Kn}$ cases, along with a tendency to underpredict the temperature across the channel at high $\mathrm{Kn}$ conditions. These behaviors reflect the inherent limitation of the G13 production terms, which are formulated under near-equilibrium assumptions.

\subsection{\label{subsec:Cylinder}Hypersonic flow around a cylinder}


\begin{table}[b]
\caption{\label{table:4_Cyl-Condition}Flow conditions for hypersonic flow around a cylinder.}
\begin{ruledtabular}
\begin{tabular}{cccccccc}
    Case & Kn & $n^{\mathrm{He}} \, [\mathrm{m}^{-3}]$ & $n^{\mathrm{Ar}} \, [\mathrm{m}^{-3}]$ & $n^{\mathrm{Kr}} \, [\mathrm{m}^{-3}]$ & $T_\infty \, [\mathrm{K}]$ & $T_{\mathrm{wall}} \, [\mathrm{K}]$ & $U_\infty \, [\mathrm{m/s}]$ \\
    \hline
    D & $0.02$& $4.052 \times 10^{19}$& $6.078 \times 10^{19}$ & $1.013 \times 10^{20}$  & $200$ & $300$ & $2251.25$ \\
\end{tabular}
\end{ruledtabular}
\end{table}

A 2D hypersonic flow around a cylinder is considered to evaluate the accuracy of the UBGK model in the presence of a strong shock-wave that involves high non-equilibrium. In contrast to previous binary mixture test cases, this case employs a ternary $(\mathrm{He-Ar-Kr})$ mixture, extending the validation to three interacting species with increased mass disparity. The cylinder has a diameter of $0.3048~\mathrm{m}$ with a uniform temperature of $300\mathrm{K}$, and its surface is assumed to be fully diffusive. The working gas has a fixed mole fractions of $\left( \chi^{\mathrm{He}}, \chi^{\mathrm{Ar}}, \chi^{\mathrm{Kr}} \right) = \left( 0.2, 0.3, 0.5 \right)$, and the simulation is conducted for a Knudsen number of $\mathrm{Kn}=0.02$. The freestream velocity and temperature are set to $U_\infty=2251.25~\mathrm{m/s}$ and $T_\infty=200~\mathrm{K}$, which corresponds to a freestream Mach number of $\mathrm{Ma}=10$. A half-body simulation is employed by imposing a reflective boundary along the x-axis as the symmetry plane. The computational domain spans $1~\mathrm{m} \times 0.8~\mathrm{m}$, with the cylinder center located at the midpoint of the reflective boundary. The domain is discretized using a uniform $1000 \times 800$ Cartesian grid, which corresponds to a cell size of $\Delta x = \Delta y = 1/6 \cdot \lambda_{mix}$, and a time step of $\Delta t = 0.01 \cdot \tau_{mix} = 2 \times 10^{-7}~\mathrm{s}$ is employed. To ensure statistical validity, $200$ simulated particles per cell is initiated at the beginning of the simulation, and the total number of particles in the statistically stationary state is $2.1 \times 10^8$. Detailed flow conditions are summarized in Table \ref{table:4_Cyl-Condition}.


\begin{figure*}[t]
	\centering
	\begin{subfigure}{0.485\textwidth}
		\centering
        \includegraphics[width=1.\linewidth]{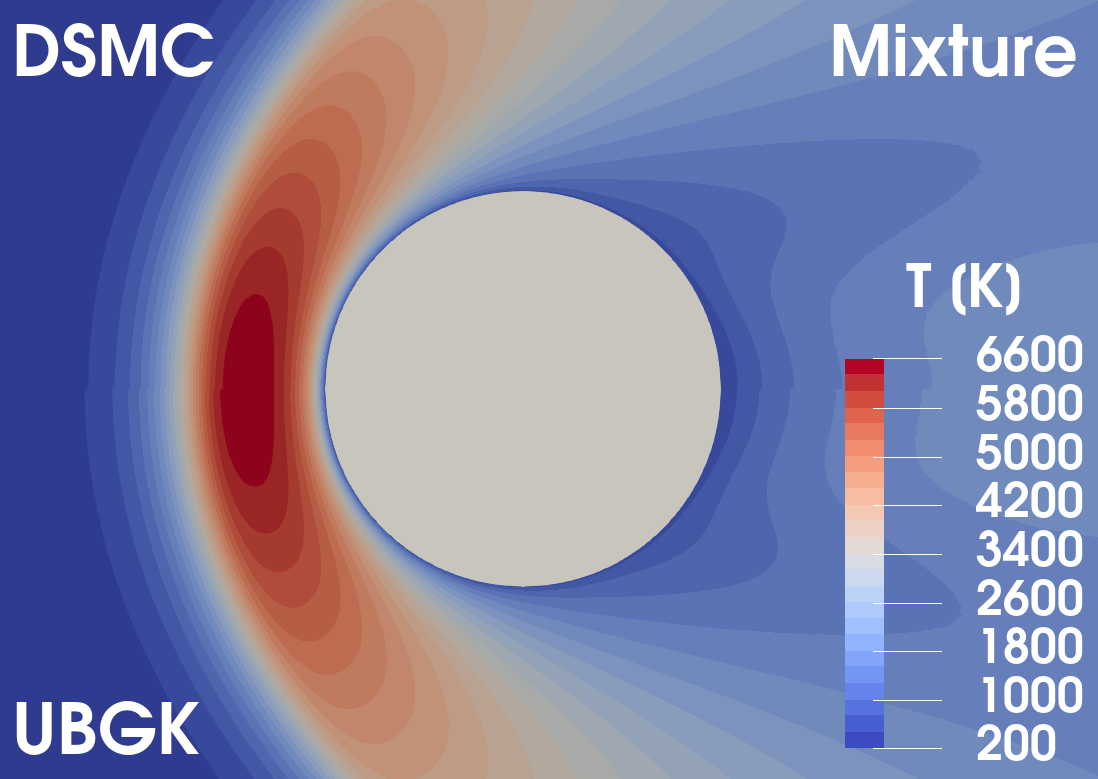}
            \caption{\label{subfig:4_Cyl-Temp-Mix}Mixture temperature.}
	\end{subfigure}
	\begin{subfigure}{0.485\textwidth}
		\centering
        \includegraphics[width=1.\linewidth]{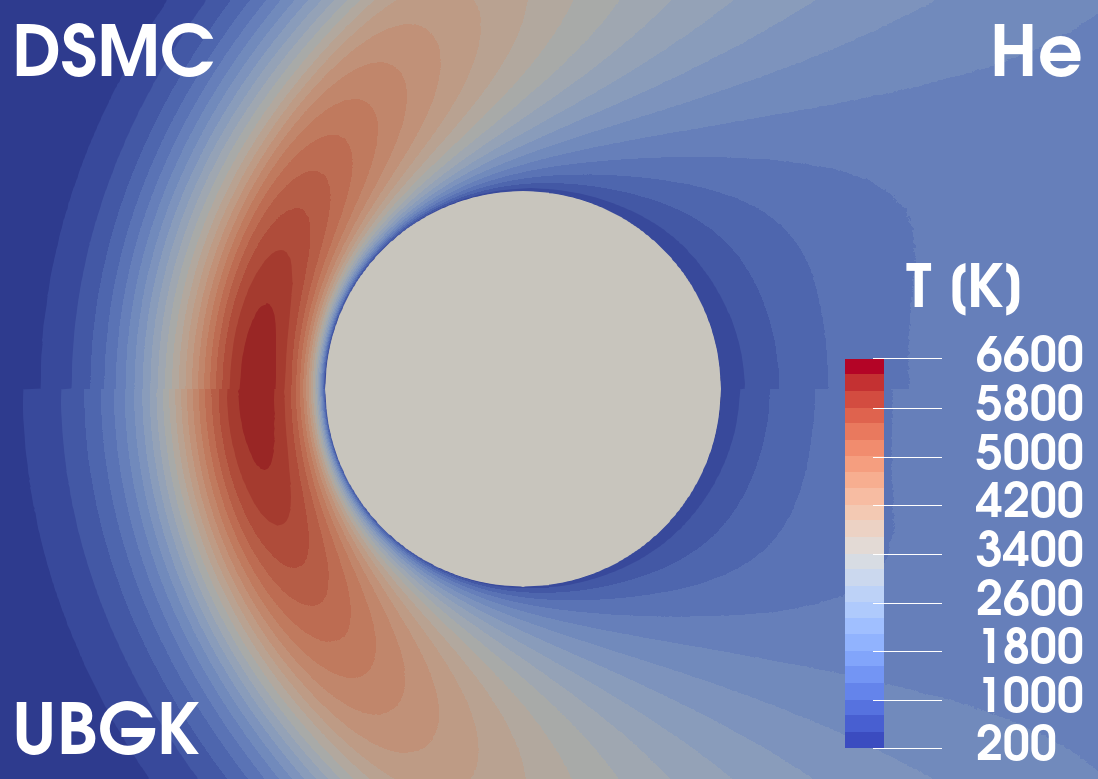}
            \caption{\label{subfig:4_Cyl-Temp-He}Helium temperature.}
	\end{subfigure}
	\centering
	\begin{subfigure}{0.485\textwidth}
		\centering
        \includegraphics[width=1.\linewidth]{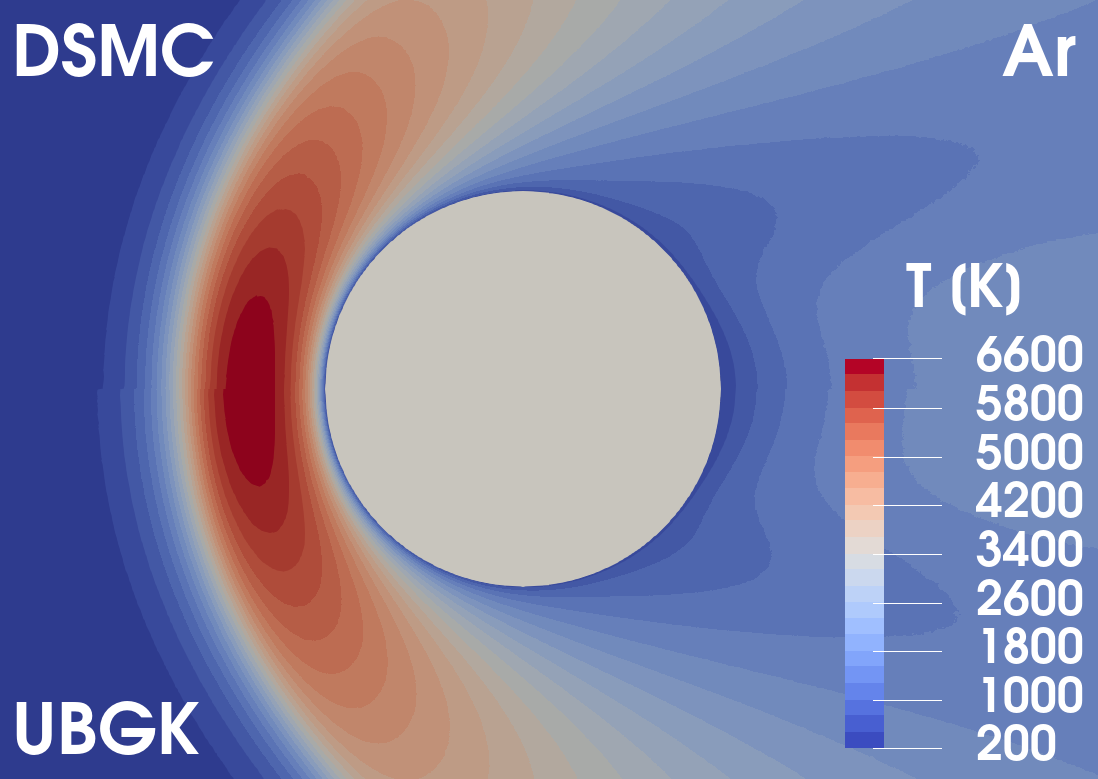}
            \caption{\label{subfig:4_Cyl-Temp-Ar}Argon temperature.}
	\end{subfigure}
	\begin{subfigure}{0.485\textwidth}
		\centering
        \includegraphics[width=1.\linewidth]{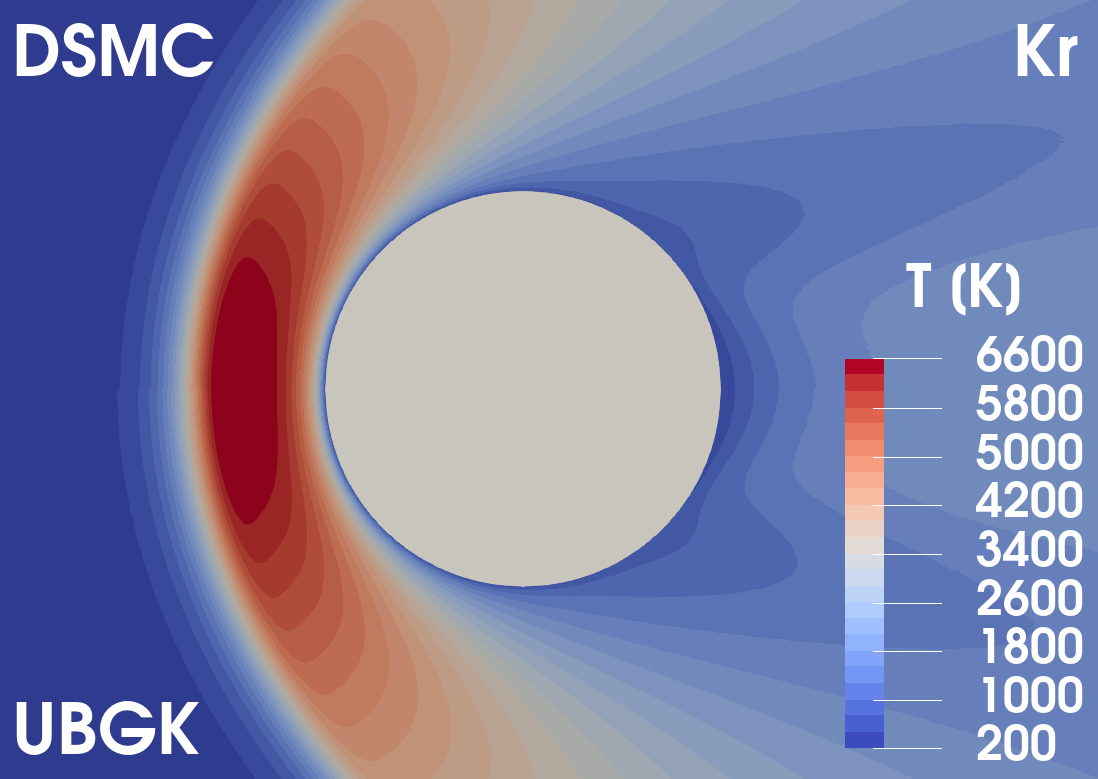}
            \caption{\label{subfig:4_Cyl-Temp-Kr}Krypton temperature.}
	\end{subfigure}
	\caption{\label{fig:4_Cyl-Temp}Temperature contours for the mixture, Helium, Argon, and Krypton of hypersonic flow around a cylinder.}
\end{figure*}

Figure \ref{fig:4_Cyl-Temp} presents $T^\alpha$ contours obtained from DSMC and the UBGK model for the mixture and individual species. The upper half represents DSMC results, and the lower half represents the UBGK predictions. Considering overall $T^\alpha$ fields of the mixture and individual species, distinct features in terms of shock thickness and peak temperature can be observed, reflecting strong non-equilibrium effects within the shock layer. While the UBGK model reproduces the overall temperature distribution with good level of agreement, a noticeable discrepancy in shock thickness is observed for Helium.


\begin{figure}[t]
    \centering
	\begin{subfigure}{0.325\textwidth}
            \centering
            \includegraphics[width=1.0\linewidth]{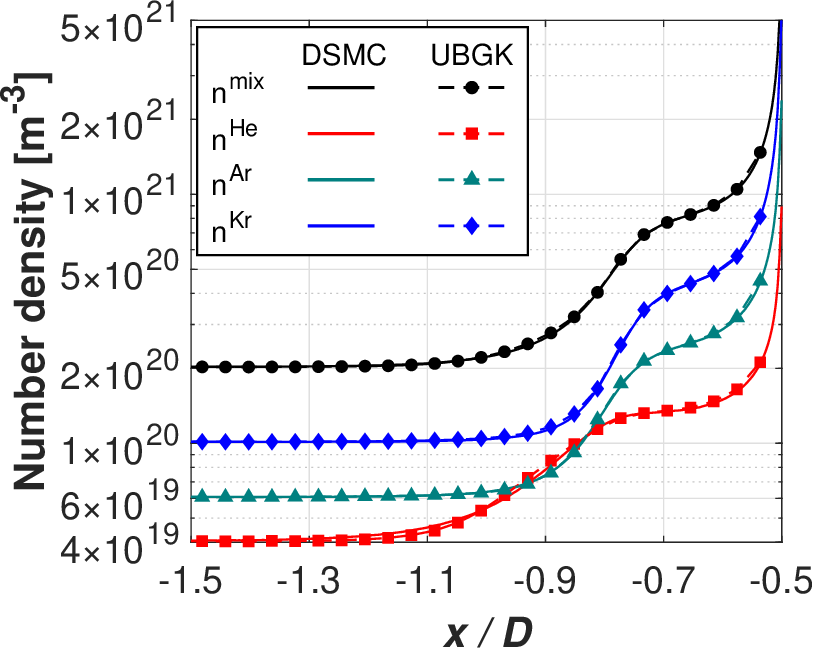}
            \caption{\label{subfig:4_Cyl-StagTempVeloNRho-NRho}Number density.}
	\end{subfigure}
    \centering
	\begin{subfigure}{0.325\textwidth}
            \centering
            \includegraphics[width=1.0\linewidth]{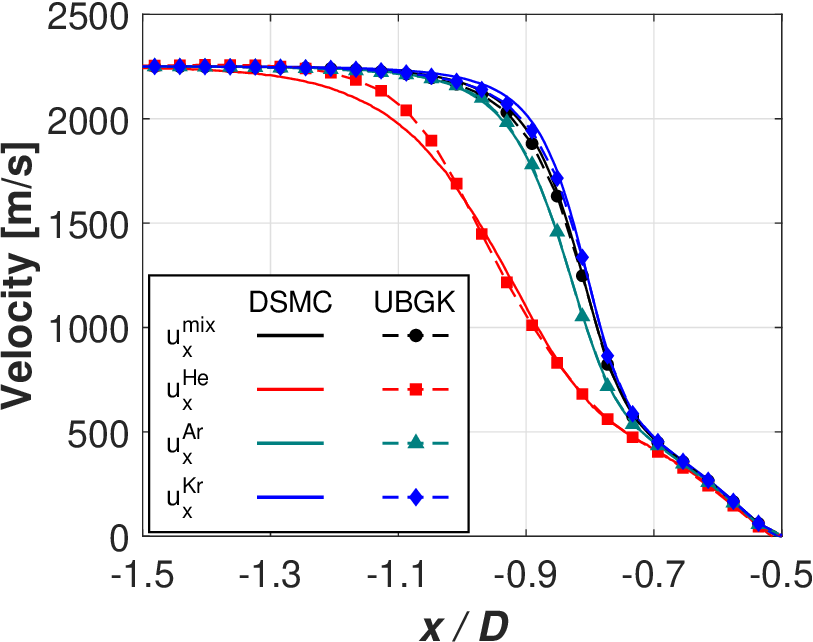}
            \caption{\label{subfig:4_Cyl-StagTempVeloNRho-Velo}Velocity.}
	\end{subfigure}
    \centering
	\begin{subfigure}{0.325\textwidth}
            \centering
            \includegraphics[width=1.0\linewidth]{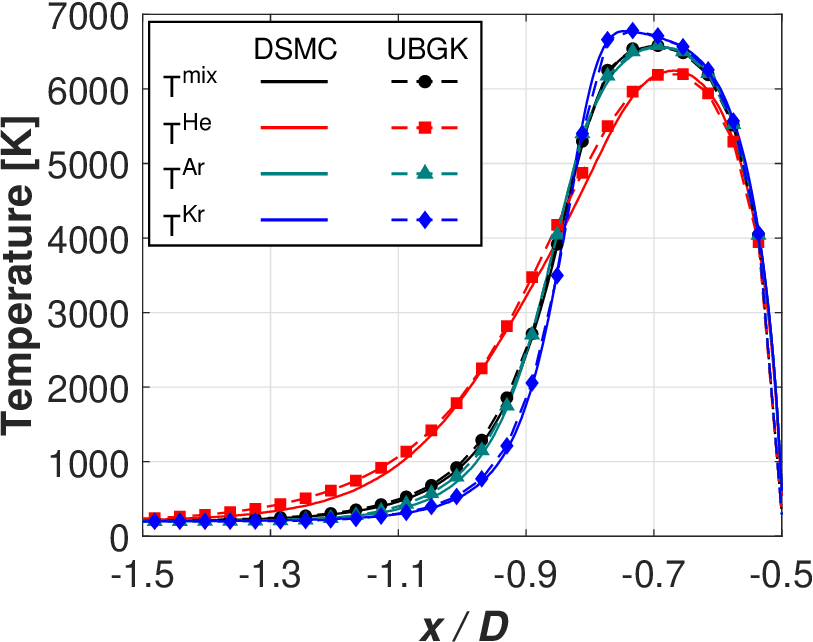}
            \caption{\label{subfig:4_Cyl-StagTempVeloNRho-StagTemp}Temperature.}
	\end{subfigure}
	\caption{\label{fig:4_Cyl-StagTempVeloNRho}Number density, velocity, and temperature distribution along the stagnation line of hypersonic flow around a cylinder.}
\end{figure} 

Figure \ref{fig:4_Cyl-StagTempVeloNRho} compares $n^\alpha$, $u_x^\alpha$, and $T^\alpha$ profiles along the stagnation line obtained from the UBGK model and DSMC. Figure \ref{subfig:4_Cyl-StagTempVeloNRho-NRho} illustrates $n^\alpha$ distributions, which show two distinct increases along the stagnation line. The first rise occurs across the shock-wave due to compression, while the second occurs near the surface as flow decelerates and accumulates towards the stagnation point. Figure \ref{subfig:4_Cyl-StagTempVeloNRho-Velo} illustrates $u_x^\alpha$ profiles, showing that $u_x^\alpha$ decreases toward zero as the flow approaches the wall. Figure \ref{subfig:4_Cyl-StagTempVeloNRho-StagTemp} illustrates $T^\alpha$ profiles, demonstrating temperature rise due to compressive heating. Across all quantities, Helium exhibits earlier changes compared to Argon and Krypton. Specifically, $n^\alpha$ rises earlier across the shock, $u_x^\alpha$ decreases earlier toward the wall, and $T^\alpha$ increases earlier within the shock layer. This behavior is attributed to the lower molecular mass of Helium, which results in higher thermal velocity and faster diffusion across the shock layer.\cite{Grad1960-ShockSpeciesSeparation, Li2024-MRMixBGK} The UBGK model agrees well with DSMC for the trends in $T^\alpha$, $n^\alpha$ and $u_x^\alpha$ along the stagnation line. However, noticeable differences between the UBGK model predictions and DSMC can be observed. These discrepancies can be attributed to the inherent formulation of the UBGK model, which combines the ESBGK and SBGK formulations. As reported by Fei \textit{et al.}\cite{Fei2020-BGKShockwave} and Park \textit{et al.}\cite{Park2024-BGKCompare}, the ESBGK model tends to overpredict $T^\alpha$, while both ESBGK and SBGK models show deviations in $n^\alpha$ and $u_x^\alpha$. The present results reflect a partial inheritance of ESBGK and SBGK model-related shock characteristics.


\begin{figure}[t]
    \centering
	\begin{subfigure}{0.495\textwidth}
		\centering
        \includegraphics[width=1.0\linewidth]{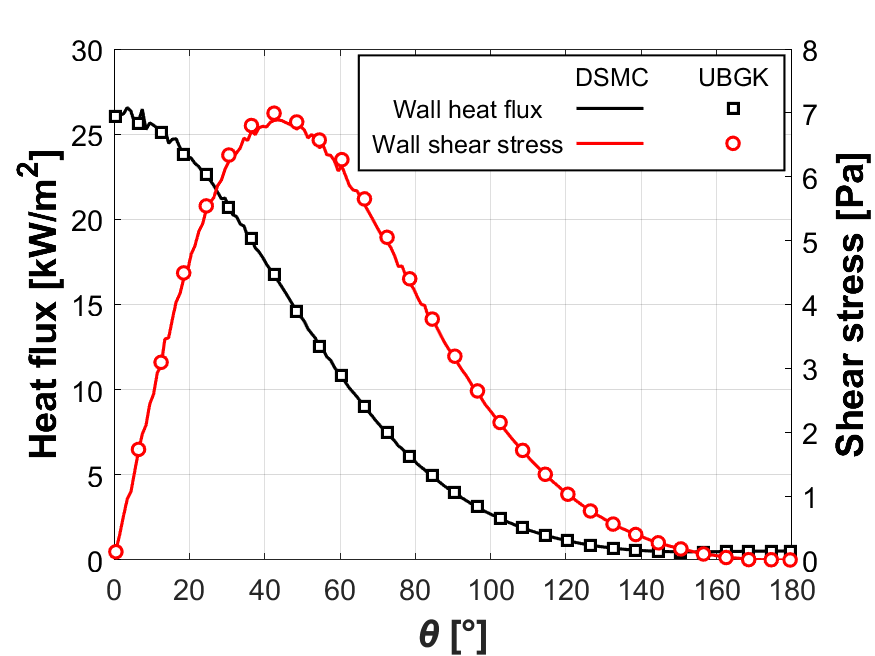}
	\end{subfigure}
    \caption{\label{fig:4_Cyl-HeatShear}Wall heat flux and shear stress distribution over a cylinder.}
\end{figure}

\begin{table}[b]
\caption{\label{table:4_Cyl-HeatDragPerError}Comparison of peak wall heat flux, total drag, and computation time between DSMC and the UBGK model. The percentage errors of the UBGK model relative to DSMC are shown in parentheses.}
\begin{ruledtabular}
\begin{tabular}{l cccc}
    Method & Peak wall heat flux $ \, [\mathrm{W/m^2}]$ & Total drag $ \, [\mathrm{N/m}]$ & Computation time $[\mathrm{s}]$ \\
    \hline
    DSMC & $26102$ & $9.8647$ & $43719$ \\
    UBGK & $26066 \, (0.1379~\%)$ & $9.8941 \, (0.2980~\%)$ & $49360 \, (12.90~\%)$ \\
\end{tabular}
\end{ruledtabular}
\end{table}

Figure \ref{fig:4_Cyl-HeatShear} compares $q^{mix}$ and $\sigma^{mix}$ distributions along the surface of the cylinder, and Table \ref{table:4_Cyl-HeatDragPerError} summarizes peak wall heat flux and total drag obtained from DSMC and the UBGK model. In Fig. \ref{fig:4_Cyl-HeatShear}, $q^{mix}$ peaks at the stagnation point and decreases toward the rear, while $\sigma^{mix}$ increases from zero, reaches a maximum, and decays downstream. Overall, the UBGK model shows good agreement with DSMC in capturing these surface transport characteristics. From Table \ref{table:4_Cyl-HeatDragPerError}, peak heat flux and total drag predicted by the UBGK model and DSMC are compared. The UBGK model predicts a peak heat flux of $26066~\mathrm{W/m^2}$ compared to $26102~\mathrm{W/m^2}$ from DSMC, and a total drag of $9.8941~\mathrm{N/m}$ compared to $9.8647~\mathrm{N/m}$ from DSMC, corresponding to relative differences below $0.14~\%$ and $0.3~\%$. The computational cost is also summarized in the last column of Table \ref{table:4_Cyl-HeatDragPerError}. The computation times were obtained using 192 cores on an AMD EPYC 9654 processor with a clock speed of $2.4~\mathrm{GHz}$. Despite the comparable accuracy, the computational cost of the UBGK model turned out to be $12.90~\%$ higher than that of DSMC. This increase is primarily attributed to the additional cell-wise macroscopic property evaluations, as well as the first-order accurate scheme of current numerical implementation. Extension to a higher-order accurate scheme is expected to alleviate this limitation. A detailed investigation of computational efficiency with respect to time step size and cell size is presented in the following section.

\subsection{\label{subsec:ComputationCostComparison}Computational cost comparison}


\begin{figure*}[t]
    \centering
    \begin{subfigure}{0.485\textwidth}
        \centering
        \includegraphics[width=1.\linewidth]{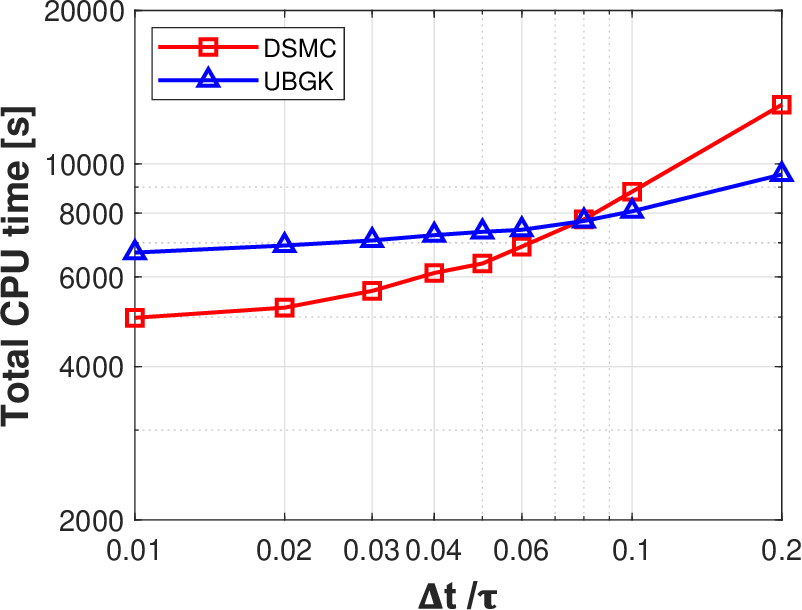}
            \caption{\label{subfig:5_Cost-Total}Total CPU time comparison.}
    \end{subfigure}
    \begin{subfigure}{0.485\textwidth}
        \centering
        \includegraphics[width=1.\linewidth]{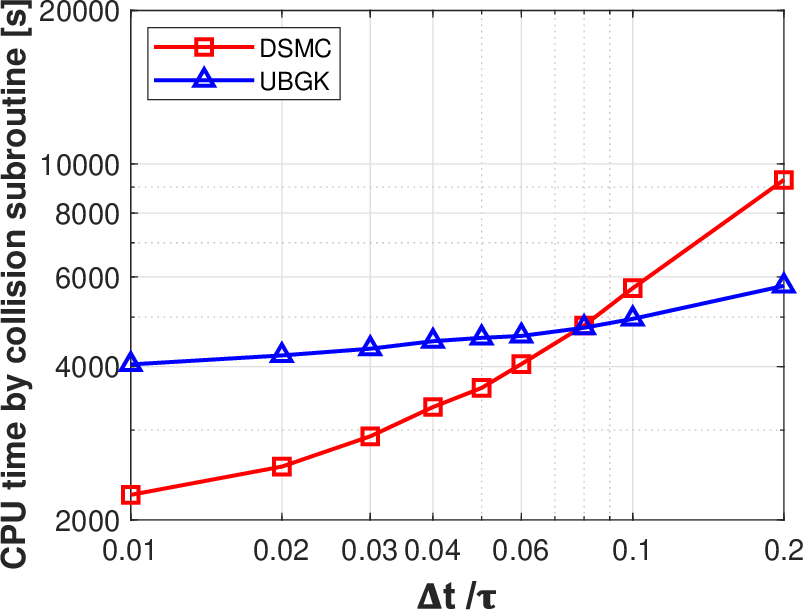}
            \caption{\label{subfig:5_Cost-Coll}Collision subroutine CPU time comparison.}
    \end{subfigure}
    \caption{\label{fig:5_Cost}Comparison of total CPU time and CPU time by collision subroutine between DSMC and the UBGK model.}
\end{figure*}

Computational cost comparison between DSMC and the UBGK model is carried out using the homogeneous relaxation case, with varying time step size $\Delta t / \tau_{mix}$. Simulations are performed using a single core on an Intel Xeon Gold 6248R processor with a clock speed of $3.0~\mathrm{GHz}$. Figure \ref{subfig:5_Cost-Total} presents the total CPU time, while Fig. \ref{subfig:5_Cost-Coll} shows the CPU time associated with the collision and relaxation subroutine for DSMC and the UBGK model. Both figures exhibit similar trends. At small $\Delta t / \tau_{mix}$, DSMC is more efficient. As $\Delta t / \tau_{mix}$ increases, the computational cost of DSMC increases more rapidly with $\Delta t / \tau_{mix}$, with a crossover occurring at $\Delta t/\tau_{mix} \approx 0.08$. This implies that the UBGK model can be advantageous for simulations that require large time step sizes.


\begin{table}[t]
\caption{\label{table:Couette-accuracy}Numerical conditions for Couette flow.}
\begin{ruledtabular}
\begin{tabular}{l cc}
    Case & $\Delta x$ & $\Delta t$ \\
    \hline
    Reference & $1 \times 10^{-3}~\mathrm{m}~(\Delta x_{ref})$ & $1 \times 10^{-6}~\mathrm{s}~(\Delta t_{ref})$ \\
    E-1 &           $\Delta x_{ref}$ &  $4 \times \Delta t_{ref}$ \\
    E-2 &           $\Delta x_{ref}$ & $10 \times \Delta t_{ref}$ \\
    E-3 &           $\Delta x_{ref}$ & $20 \times \Delta t_{ref}$ \\
    E-4 &           $\Delta x_{ref}$ & $40 \times \Delta t_{ref}$ \\
    E-5 &  $4 \times \Delta x_{ref}$ &           $\Delta t_{ref}$ \\
    E-6 & $10 \times \Delta x_{ref}$ &           $\Delta t_{ref}$ \\
    E-7 & $20 \times \Delta x_{ref}$ &           $\Delta t_{ref}$ \\
    E-8 & $40 \times \Delta x_{ref}$ &           $\Delta t_{ref}$ \\
\end{tabular}
\end{ruledtabular}
\end{table}

\begin{figure*}[t]
    \centering
    \begin{subfigure}{0.485\textwidth}
        \centering
        \includegraphics[width=1.\linewidth]{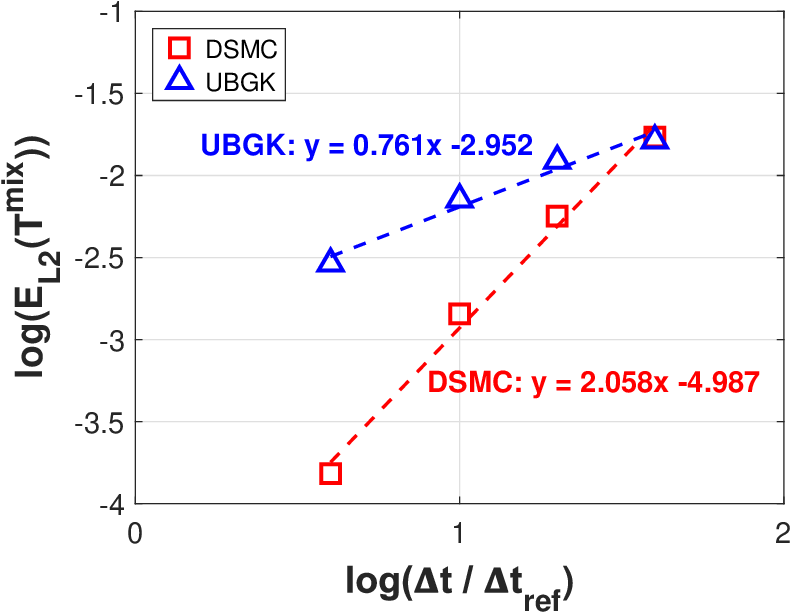}
            \caption{\label{subfig:5_TS-accuracy}Relative L2-norm error with respect to time step sizes.}
    \end{subfigure}
    \begin{subfigure}{0.485\textwidth}
        \centering
        \includegraphics[width=1.\linewidth]{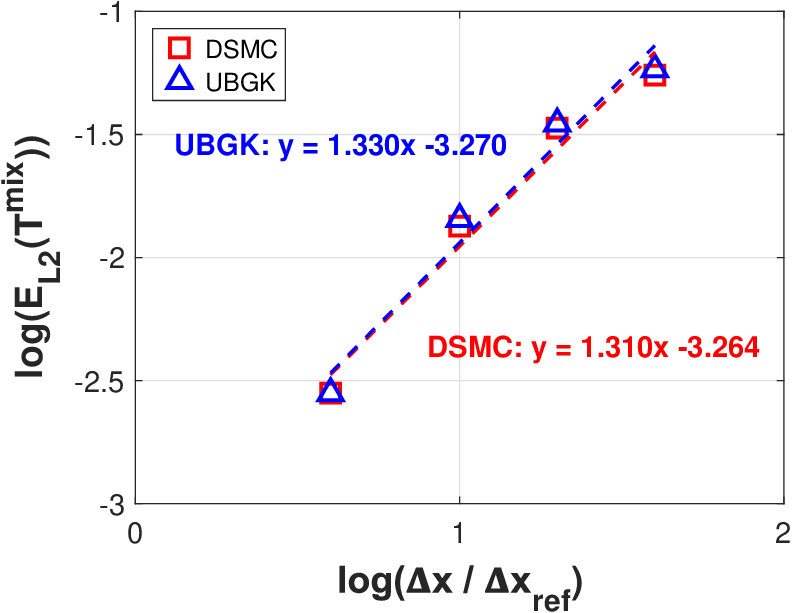}
            \caption{\label{subfig:5_CS-accuracy}Relative L2-norm error with respect to cell sizes.}
    \end{subfigure}
    \caption{\label{fig:5_Couette-accuracy}Relative L2-norm errors of temperature of the mixture in Couette flow.}
\end{figure*}

The temporal and spatial accuracy of the current UBGK scheme is investigated using the Couette flow case. The relative $\mathrm{L}2$-norm error of the mixture temperature, i.e. $T^{mix}$, obtained from the UBGK model is evaluated with respect to DSMC for different temporal and spatial discretizations. The Knudsen number is fixed to $0.005$, and the flow condition is identical to the case $\mathrm{C}$-1 of Couette flow from Section \ref{subsec:Couette}. The numerical conditions are summarized in Table \ref{table:Couette-accuracy}. Cases (E-1)$-$(E-4) evaluate the temporal accuracy by varying the time step as $\Delta t / \Delta t_{ref} \in \{4, 10, 20, 40\}$ under a fixed cell size, and cases (E-5)$-$(E-8) evaluate the spatial accuracy by varying the cell size as $\Delta x / \Delta x_{ref} \in \{4, 10, 20, 40\}$ under a fixed time step. Figure \ref{fig:5_Couette-accuracy} shows the variation of relative $\mathrm{L}2$-norm errors with respect to time step sizes and cell sizes. Figure \ref{subfig:5_TS-accuracy} demonstrates that, when the time step size is varied, the evaluated orders of convergence of DSMC and the UBGK model are $2.058$ and $0.761$, respectively. Figure \ref{subfig:5_CS-accuracy} shows that, when the cell size is varied, the evaluated orders of convergence are $1.310$ and $1.330$. The temporal convergence of DSMC is observed to be second-order, while DSMC remains first-order accurate in both time and space due to the de-coupled treatment of particle motion and collisions. It has been reported that the error profile in DSMC exhibits a steeper slope at small $\Delta t$ and a milder slope at larger $\Delta t$.\cite{Kim2025-FPInterpolation} In the present study, the reference time step is set to $\Delta t_{ref} = 0.1~\tau_c$, and the evaluated range falls within the small-$\Delta t$ regime, where the slope is steeper. Under the same de-coupled framework, the UBGK scheme shows first-order convergence in both time and space. This is because particle advection and relaxation are treated in a de-coupled manner, with macroscopic properties assumed constant within each cell over a time step.\cite{Fei2020-USPBGK, Pfeiffer2025-CNBGK, Kim2025-FPInterpolation} To improve computational efficiency while maintaining accuracy, extension to higher-order schemes is necessary.

\section{\label{sec:conclusions}Conclusion}


In this work, a mixture UBGK model for rarefied monatomic gas mixtures is developed based on the multi-relaxation BGK formulation to reproduce correct NSF-level transport behavior in the near-continuum regime. The single-species unified BGK (UBGK) model is extended to gas mixtures, and the relaxation properties are determined by matching the production terms of the mixture UBGK equation to those of the Boltzmann equation. The developed mixture UBGK model was implemented within the particle framework and validated against DSMC using four benchmark problems. Homogeneous relaxation demonstrates that the temporal evolution of macroscopic properties predicted by the mixture UBGK model closely matches DSMC reference solutions, while Poiseuille and Couette flows confirm that it reproduces mixture and species property distributions over a range of mole fractions and Knudsen numbers. Finally, a hypersonic flow around a cylinder demonstrates that the mixture UBGK model captures the shock structure, peak surface heat flux, and aerodynamic drag in good agreement with DSMC. In terms of computational performance, cost analysis of homogeneous relaxation shows that the mixture UBGK model becomes more efficient than DSMC at sufficiently large time step sizes. However, accuracy analysis of Couette flow confirms that the present UBGK scheme exhibits first-order convergence in both time and space, suggesting that higher-order schemes may improve computational efficiency while preserving accuracy. For future work, the current monatomic gas framework will be extended to diatomic and polyatomic gases by incorporating internal energy modes.\cite{Mathiaud2022-DiatomicBGK, Pfeiffer2019-PolyatomicBGK} In addition, the present first-order numerical scheme will be improved to higher-order accuracy using the USP-BGK framework.\cite{Fei2020-USPBGK} Furthermore, coupling of the mixture UBGK model with hybrid DSMC–UBGK approaches will be explored to enable efficient multi-scale simulations.

\begin{acknowledgments}

This work was supported by the National Research Foundation of Korea(NRF) grant funded by the Korea government(MSIT) (No. RS-2025-02213804). This work was also supported by the National Supercomputing Center with supercomputing resources including technical support(KSC-2025-CRE-0573).

\end{acknowledgments}


\section*{Data Availability}
The data that support the findings of this study are available from the corresponding author upon reasonable request.

\appendix

\section{\label{appendix:UBGKPT}Production terms of the UBGK model}


In this appendix, a detailed derivation of the UBGK production terms are presented. Substituting the UBGK model's target distribution, Eq. (\ref{eqn:UBGK-TargetDistribution}) into Eq. (\ref{eqn:BGK PT-derivation}) gives:
\begin{equation}\label{eqn:BGK production terms appendix}
\begin{aligned}
    P_{UBGK}^{\alpha\beta} \left( \psi_i^\alpha \right) 
    &= \int_{-\infty}^{\infty} m^\alpha \psi_i^\alpha \nu_{UBGK}^{\alpha\beta} \left( f_U^{\alpha\beta} - f^\alpha \right) d\boldsymbol{c}^\alpha \\
    &= m^\alpha \nu_{UBGK}^{\alpha\beta} \int_{-\infty}^{\infty} \psi_i^\alpha \left( f_U^{\alpha\beta} - f^\alpha \right) d\boldsymbol{c}^\alpha \\
    &= m^\alpha \nu_{UBGK}^{\alpha\beta} \left( \Bigl\langle \psi_i^\alpha ~|~ f_U^{\alpha\beta} \Bigl\rangle - \Bigl\langle \psi_i^\alpha ~|~ f^\alpha \Bigl\rangle \right),
\end{aligned}
\end{equation}
where the angle bracket operator $\langle \psi_i^\alpha~|~\cdot~\rangle$ is defined as:
\begin{equation}\label{eqn:UBGK moment for U}
\begin{aligned}
    \Bigl\langle \psi_i^\alpha ~|~ f_U^{\alpha\beta} \Bigl\rangle ~= \int_{- \infty}^{\infty} \psi_i^\alpha f_U^{\alpha\beta} d\boldsymbol{c}^\alpha,
\end{aligned}
\end{equation}
and
\begin{equation}\label{eqn:UBGK moment for f}
\begin{aligned}
    \Bigl\langle \psi_i^\alpha ~|~ f^\alpha \Bigl\rangle ~= \int_{- \infty}^{\infty} \psi_i^\alpha f^\alpha d\boldsymbol{c}^\alpha.
\end{aligned}
\end{equation}

The moments for Eq. (\ref{eqn:UBGK moment for U}) are computed as:
\begin{equation}\label{eqn:UBGK moment for U mass}
\begin{aligned}
    \Bigl\langle 1 ~|~ f_U^{\alpha\beta} \Bigl\rangle
    &= n^\alpha,
\end{aligned}
\end{equation}
\begin{equation}\label{eqn:UBGK moment for U momentum}
\begin{aligned}
    \Bigl\langle \hat{C}_i^\alpha ~|~ f_U^{\alpha\beta} \Bigl\rangle
    &= \Bigl\langle C_i^{\alpha\beta} + \Delta\tilde{u}_i^{\alpha\beta} ~|~ f_U^{\alpha\beta} \Bigl\rangle \\
    &= 0 + n^\alpha \Delta\tilde{u}_i^{\alpha\beta} \\
    &= n^\alpha \Delta\tilde{u}_i^{\alpha\beta},
\end{aligned}
\end{equation}
\begin{equation}\label{eqn:UBGK moment for U energy}
\begin{aligned}
    \Bigl\langle \hat{C}_j^\alpha \hat{C}_j^\alpha ~|~ f_U^{\alpha\beta} \Bigl\rangle
    &= \Bigl\langle \left( C_j^{\alpha\beta} + \Delta\tilde{u}_j^{\alpha\beta} \right) \left( C_j^{\alpha\beta} + \Delta\tilde{u}_j^{\alpha\beta} \right) ~|~ f_U^{\alpha\beta} \Bigl\rangle \\
    &= \Bigl\langle C_j^{\alpha\beta} C_j^{\alpha\beta} + 2 C_j^{\alpha\beta} \Delta\tilde{u}_j^{\alpha\beta} + \Delta\tilde{u}_j^{\alpha\beta} \Delta\tilde{u}_j^{\alpha\beta} ~|~ f_U^{\alpha\beta} \Bigl\rangle \\
    &= 3 n^\alpha R^\alpha T^{\alpha\beta} + 0 + n^\alpha \Delta\tilde{u}_j^{\alpha\beta} \Delta\tilde{u}_j^{\alpha\beta} \\
    &= 3 n^\alpha R^\alpha T^{\alpha\beta} + n^\alpha \Delta\tilde{u}_j^{\alpha\beta} \Delta\tilde{u}_j^{\alpha\beta},
\end{aligned}
\end{equation}
\begin{equation}\label{eqn:UBGK moment for U shearstress}
\begin{aligned}
    \Bigl\langle \hat{C}_{\langle i}^\alpha \hat{C}_{j \rangle}^\alpha ~|~ f_U^{\alpha\beta} \Bigl\rangle
    &= \Bigl\langle \left( C_{\langle i}^{\alpha\beta} + \Delta\tilde{u}_{\langle i}^{\alpha\beta} \right) \left( C_{j \rangle}^{\alpha\beta} + \Delta\tilde{u}_{j \rangle}^{\alpha\beta} \right) ~|~ f_U^{\alpha\beta} \Bigl\rangle \\
    &= \Bigl\langle C_{\langle i}^{\alpha\beta} C_{j \rangle}^{\alpha\beta} + C_{\langle i}^{\alpha\beta} \Delta\tilde{u}_{j \rangle}^{\alpha\beta} + \Delta\tilde{u}_{\langle i}^{\alpha\beta} C_{j \rangle}^{\alpha\beta} + \Delta\tilde{u}_{\langle i}^{\alpha\beta} \Delta\tilde{u}_{j \rangle}^{\alpha\beta} ~|~ f_U^{\alpha\beta} \Bigl\rangle \\
    &= n^\alpha C_{ES} \sigma_{ij}^{\alpha\beta} / m^\alpha + 0 + 0 + n^\alpha \Delta\tilde{u}_{\langle i}^{\alpha\beta} \Delta\tilde{u}_{j \rangle}^{\alpha\beta} \\
    &= n^\alpha C_{ES} \sigma_{ij}^{\alpha\beta} / m^\alpha + n^\alpha \Delta\tilde{u}_{\langle i}^{\alpha\beta} \Delta\tilde{u}_{j \rangle}^{\alpha\beta},
\end{aligned}
\end{equation}
\begin{equation}\label{eqn:UBGK moment for U heatflux}
\begin{aligned}
    \Bigl\langle \hat{C}_i^\alpha \hat{C}_j^\alpha \hat{C}_j^\alpha ~|~ f_U^{\alpha\beta} \Bigl\rangle
    &= \Bigl\langle \left( C_i^{\alpha\beta} + \Delta\tilde{u}_i^{\alpha\beta} \right) \left( C_j^{\alpha\beta} + \Delta\tilde{u}_j^{\alpha\beta} \right) \left( C_j^{\alpha\beta} + \Delta\tilde{u}_j^{\alpha\beta} \right) ~|~ f_U^{\alpha\beta} \Bigl\rangle \\
    &= \Bigl\langle C_i^{\alpha\beta} C_j^{\alpha\beta} C_j^{\alpha\beta} + 2 C_i^{\alpha\beta} C_j^{\alpha\beta} \Delta\tilde{u}_j^{\alpha\beta} + \Delta\tilde{u}_i^{\alpha\beta} C_j^{\alpha\beta} C_j^{\alpha\beta} \\
    &\quad\quad + 2 \Delta\tilde{u}_i^{\alpha\beta} C_j^{\alpha\beta} \Delta\tilde{u}_j^{\alpha\beta} + C_i^{\alpha\beta} \Delta\tilde{u}_j^{\alpha\beta} \Delta\tilde{u}_j^{\alpha\beta} + \Delta\tilde{u}_i^{\alpha\beta} \Delta\tilde{u}_j^{\alpha\beta} \Delta\tilde{u}_j^{\alpha\beta} ~|~ f_U^{\alpha\beta} \Bigl\rangle \\
    &= 2 C_S q_i^{\alpha\beta} / m^\alpha + 2 \left( C_{ES} \sigma_{ij}^{\alpha\beta} / m^\alpha + n^\alpha R^\alpha T^{\alpha\beta} \delta_{ij} \right) \Delta\tilde{u}_j^{\alpha\beta} \\
    &\quad\quad + \Delta\tilde{u}_i^{\alpha\beta} \left( 3 n^\alpha R^\alpha T^{\alpha\beta} \right) + 0 + 0 + n^{\alpha\beta} \Delta\tilde{u}_i^{\alpha\beta} \Delta\tilde{u}_j^{\alpha\beta} \Delta\tilde{u}_j^{\alpha\beta} \\
    &= 2 C_S q_i^{\alpha\beta} / m^\alpha + 2 C_{ES} \sigma_{ij}^{\alpha\beta} \Delta\tilde{u}_j^{\alpha\beta} / m^\alpha + 5 n^\alpha R^\alpha T^{\alpha\beta} \Delta\tilde{u}_i^{\alpha\beta} \\
    &\quad\quad + n^\alpha \Delta\tilde{u}_i^{\alpha\beta} \Delta\tilde{u}_j^{\alpha\beta} \Delta\tilde{u}_j^{\alpha\beta},
\end{aligned}
\end{equation}
where $\Delta\tilde{u}_i^{\alpha\beta} = \hat{u}_i^\alpha + u_i^{\alpha\beta} - u_i^\alpha$. Since the BGK models are constructed from a Maxwellian ansatz, the deviatoric part of the even-order moments, and the odd-order moments of the thermal velocity would vanish. However, the deviatoric part of the second-order moment, i.e. $\langle C_{\langle i}^\alpha C_{j \rangle}^\alpha | f_U^{\alpha\beta} \rangle$, which is related to shear stress, has a non-zero value, since it is fixed by the ESBGK component. Similarly, the third-order moment, i.e. $\langle C_i^\alpha C_j^\alpha C_j^\alpha | f_U^{\alpha\beta} \rangle$, which is related to heat flux, also has a non-zero value, since it is fixed by the SBGK component.

Next, the moments for Eq. (\ref{eqn:UBGK moment for f}) are calculated as:
\begin{equation}\label{eqn:UBGK moment for f mass}
\begin{aligned}
    \Bigl\langle 1 ~|~ f^\alpha \Bigl\rangle
    &= n^\alpha,
\end{aligned}
\end{equation}
\begin{equation}\label{eqn:UBGK moment for f momentum}
\begin{aligned}
    \Bigl\langle \hat{C}_i^\alpha ~|~ f^\alpha \Bigl\rangle
    &= \Bigl\langle C_i^\alpha + \hat{u}_i^\alpha ~|~ f^\alpha \Bigl\rangle \\
    &= 0 + n^\alpha \hat{u}_i^\alpha \\
    &= n^\alpha \hat{u}_i^\alpha,
\end{aligned}
\end{equation}
\begin{equation}\label{eqn:UBGK moment for f energy}
\begin{aligned}
    \Bigl\langle \hat{C}_j^\alpha \hat{C}_j^\alpha ~|~ f^\alpha \Bigl\rangle
    &= \Bigl\langle \left( C_j^\alpha + \hat{u}_j^\alpha \right) \left( C_j^\alpha + \hat{u}_j^\alpha \right) ~|~ f^\alpha \Bigl\rangle \\
    &= \Bigl\langle C_j^\alpha C_j^\alpha + 2 C_j^\alpha \hat{u}_j^\alpha + \hat{u}_j^\alpha \hat{u}_j^\alpha ~|~ f^\alpha \Bigl\rangle \\
    &= 3 n^\alpha R^\alpha T^\alpha + 0 + n^\alpha \hat{u}_j^\alpha \hat{u}_j^\alpha \\
    &= 3 n^\alpha R^\alpha T^\alpha + n^\alpha \hat{u}_j^\alpha \hat{u}_j^\alpha,
\end{aligned}
\end{equation}
\begin{equation}\label{eqn:UBGK moment for f sherstress}
\begin{aligned}
    \Bigl\langle \hat{C}_{\langle i}^\alpha \hat{C}_{j \rangle}^\alpha ~|~ f^\alpha \Bigl\rangle
    &= \Bigl\langle \left( C_{\langle i}^\alpha + \hat{u}_{\langle i}^\alpha \right) \left( C_{j \rangle}^\alpha + \hat{u}_{j \rangle}^\alpha \right) ~|~ f^\alpha \Bigl\rangle \\
    &= \Bigl\langle C_{\langle i}^\alpha C_{j \rangle}^\alpha + C_{\langle i}^\alpha \hat{u}_{j \rangle}^\alpha + \hat{u}_{\langle i}^\alpha C_{j \rangle}^\alpha + \hat{u}_{\langle i}^\alpha \hat{u}_{j \rangle}^\alpha ~|~ f^\alpha \Bigl\rangle \\
    &= \sigma_{ij}^\alpha / m^\alpha + 0 + 0 + n^\alpha \hat{u}_{\langle i}^\alpha \hat{u}_{j \rangle}^\alpha \\
    &= \sigma_{ij}^\alpha / m^\alpha + n^\alpha \hat{u}_{\langle i}^\alpha \hat{u}_{j \rangle}^\alpha,
\end{aligned}
\end{equation}
\begin{equation}\label{eqn:UBGK moment for f heatflux}
\begin{aligned}
    \Bigl\langle \hat{C}_i^\alpha \hat{C}_j^\alpha \hat{C}_j^\alpha ~|~ f^\alpha \Bigl\rangle
    &= \Bigl\langle \left( C_i^\alpha + \hat{u}_i^\alpha \right) \left( C_j^\alpha + \hat{u}_j^\alpha \right) \left( C_j^\alpha + \hat{u}_j^\alpha \right) ~|~ f^\alpha \Bigl\rangle \\
    &= \Bigl\langle C_i^\alpha C_j^\alpha C_j^\alpha + 2 C_i^\alpha C_j^\alpha \hat{u}_j^\alpha + \hat{u}_i^\alpha C_j^\alpha C_j^\alpha + 2 \hat{u}_i^\alpha C_j^\alpha \hat{u}_j^\alpha + C_i^\alpha \hat{u}_j^\alpha \hat{u}_j^\alpha + \hat{u}_i^\alpha \hat{u}_j^\alpha \hat{u}_j^\alpha ~|~ f^\alpha \Bigl\rangle \\
    &= 2 q_i^\alpha / m^\alpha + 2 \left( \sigma_{ij}^{\alpha\beta} / m^\alpha + n^\alpha R^\alpha T^\alpha \delta_{ij} \right) \hat{u}_j^{\alpha\beta} + \hat{u}_i^\alpha \left( 3 n^\alpha R^\alpha T^\alpha \right) \\
    &\quad\quad + 0 + 0 + n^\alpha \hat{u}_i^\alpha \hat{u}_j^\alpha \hat{u}_j^\alpha \\
    &= 2 q_i^\alpha / m^\alpha + 2 \sigma_{ij}^{\alpha\beta} \hat{u}_j^{\alpha\beta} / m^\alpha + 5 n^\alpha R^\alpha T^\alpha \hat{u}_i^{\alpha\beta} + n^\alpha \hat{u}_i^\alpha \hat{u}_j^\alpha \hat{u}_j^\alpha.
\end{aligned}
\end{equation}
Combining Eqs. (\ref{eqn:UBGK moment for U mass})-(\ref{eqn:UBGK moment for U heatflux}) with Eqs. (\ref{eqn:UBGK moment for f mass})-(\ref{eqn:UBGK moment for f heatflux}) and Eq. (\ref{eqn:BGK production terms appendix}), the production terms for the UBGK model, Eqs. (\ref{eqn:UBGK PT-mass})-(\ref{eqn:UBGK PT-heatflux}) are derived.

\section{\label{appendix:T-sign-Discussion}Discussion on positivity of target temperature}


In this appendix, the conditions required to ensure the positivity of the relaxation temperature $T^{\alpha\beta}$ are discussed. Separating the relaxation temperature in Eq. (\ref{eqn:UBGK-relaxation-temperature}) into the species-specific component and the nonlinear contribution gives:
\begin{equation}\label{eqn:UBGK-relaxation-temperature-decompose}
\begin{aligned}
    T^{\alpha\beta} = T^\alpha + T_N^{\alpha\beta},
\end{aligned}
\end{equation}
the nonlinear contribution term becomes,
\begin{equation}\label{eqn:UBGK-relaxation-temperature-nonlinearcontribution}
\begin{aligned}
    T_N^{\alpha\beta} 
    &= - \frac{1}{3 R^\alpha} \Bigg[
        \Delta \tilde{u}_j^{\alpha\beta} \Delta \tilde{u}_j^{\alpha\beta} - \hat{u}_j^\alpha \hat{u}_j^\alpha \\
        &\quad +\mu^\beta \mathrm{VHS} \left[ 1 \right] \frac{\nu_{G13}^{\alpha\beta}}{\nu_{UBGK}^{\alpha\beta}} \Big[
            10 \hat{\theta}^{\alpha\beta} \Delta \hat{\theta}^{\alpha\beta} 
            -\frac{10}{3} \left( \mu^\alpha - \mu^\beta \right) \hat{u}_j^\alpha \hat{u}_j^\beta
        \Big]
    \Bigg].
\end{aligned}
\end{equation}
By definition, the species-specific temperature $T^\alpha$ remains positive, whereas the nonlinear component $T_N^{\alpha\beta}$ may become negative, potentially leading to a negative relaxation temperature $T^{\alpha\beta}$. To ensure positivity, $T_N^{\alpha\beta}$ is neglected whenever $T^{\alpha\beta} \leq 0$, hence $T^{\alpha\beta}$ is replaced by $T^\alpha$. While this strategy may lead to incorrect temperature relaxation, negative values were observed only in cells with fewer than 10 particles, typically in cells near solid surfaces with relatively small volume or in the downstream region of the blunt body. This suggests that the issue is a stochastic fluctuation inherent to low sample size statistics rather than a modeling limitation, with negligible impact on the bulk flow.

\bibliography{main}

\end{document}